\documentclass[iop,apj]{emulateapj}  

\usepackage{epsfig}
\usepackage{amsmath}
\usepackage{amssymb}
\usepackage{natbib}
\usepackage{threeparttable} 
\usepackage{enumerate}

\usepackage{epstopdf}

\newcommand{\chan}{\textit{Chandra}}
\newcommand{\swift}{\textit{Swift}}

\newcommand{\xmm}{\textit{XMM-Newton}}

\newcommand{\Msun}{\mathrm{M}_{\odot}}
\newcommand{\lum}{\mathrm{erg~s}^{-1}}
\newcommand{\flux}{\mathrm{erg~cm}^{-2}~\mathrm{s}^{-1}}
\newcommand{\fluence}{\mathrm{erg~cm}^{-2}}
\newcommand{\cnts}{\mathrm{c~s}^{-1}}
\newcommand{\mdot}{\mathrm{M_{\odot}~yr}^{-1}}
\newcommand{\mdotgs}{\mathrm{g~s}^{-1}}
\newcommand{\nh}{\mathrm{cm}^{-2}}
\newcommand{\dist}{(D/\mathrm{6.5~kpc})^{2}}

\newcommand{\xmmbron}{XMM J174457--2850.3}
\newcommand{\psr}{PSR J1023+0038}
\newcommand{\xss}{XSS J12270--4859}
\newcommand{\psrm}{PSR J1824--24521}
\newcommand{\igrbron}{IGR J18245--2452}
\newcommand{\sax}{SAX J1808.4--3658}
\newcommand{\saxrudy}{SAX J1750.8--2900}

\newcommand{\grobron}{GRO J1744--28}
\newcommand{\exoter}{EXO 1745--248}
\newcommand{\saxbron}{SAX~J1747.0--2853}

\newcommand{\xte}{XTE J1701--462}

\newcommand{\sgra}{Sgr~A$^{*}$}

\newcommand{\ascabron}{AX~J1745.6--2901}

\newcommand{\grsbron}{GRS~1741--2853}
\newcommand{\ksbron}{KS~1741--293}

\def \atel {ATel}

\slugcomment{Received 2014 June 12; accepted 2014 July 30; published 2014 ???}
\shorttitle{XMM~J174457--2850.3}
\shortauthors{Degenaar et al.}

\begin{document}

\title{The peculiar Galactic center neutron star X-ray binary XMM J174457--2850.3}

\author{N. Degenaar$^{1,}$\altaffilmark{8}, R. Wijnands$^{2}$, M.T. Reynolds$^1$, J.M. Miller$^{1}$, D.~Altamirano$^3$, J. Kennea$^{4}$, N.~Gehrels$^{5}$, D.~Haggard$^6$, and G.~Ponti$^7$ }
\affil{$^1$Department of Astronomy, University of Michigan, 500 Church Street, Ann Arbor, MI 48109, USA; degenaar@umich.edu\\
$^2$Anton Pannekoek Institute of Astronomy, University of Amsterdam, Science Park 904,1098 XH Amsterdam, The Netherlands\\
$^3$School of Physics and Astronomy, University of Southampton, Highfield, Southampton SO17 1BJ, UK\\
$^4$Department of Astronomy and Astrophysics, 525 Davey Lab, Pennsylvania State University, University Park, PA 16802, USA\\
$^5$Astrophysics Science Division, NASA Goddard Space Flight Center, Greenbelt, MD, USA\\
$^6$CIERA, Physics and Astronomy Department, Northwestern University, 2145 Sheridan Road, Evanston, IL 60208, USA\\
$^7$Max Planck Institute fur Extraterrestriche Physik, D-85748 Garching, Germany
}

\altaffiltext{8}{Hubble fellow}


\begin{abstract}

\end{abstract}
\begin{abstract} 
The recent discovery of a milli-second radio pulsar experiencing an accretion outburst similar to those seen in low mass X-ray binaries, has opened up a new opportunity to investigate the evolutionary link between these two different neutron star manifestations. The remarkable X-ray variability and hard X-ray spectrum of this object can potentially serve as a template to search for other X-ray binary/radio pulsar transitional objects. Here we demonstrate that the transient X-ray source \xmmbron\ near the Galactic center displays similar X-ray properties. We report on the detection of an energetic thermonuclear burst with an estimated duration of $\simeq$2~hr and a radiated energy output of $\simeq5\times10^{40}$~erg, which unambiguously demonstrates that the source harbors an accreting neutron star. It has a quiescent X-ray luminosity of $L_{\mathrm{X}} \simeq 5\times 10^{32}~\dist~\lum$ and exhibits occasional accretion outbursts during which it brightens to $L_{\mathrm{X}} \simeq 10^{35}-10^{36}~\dist~\lum$ for a few weeks (2--10 keV). However, the source often lingers in between outburst and quiescence at $L_{\mathrm{X}} \simeq 10^{33}-10^{34}~\dist~\lum$. This peculiar X-ray flux behavior and its relatively hard X-ray spectrum, a power law with an index of $\Gamma \simeq 1.4$, could possibly be explained in terms of the interaction between the accretion flow and the magnetic field of the neutron star. 
\end{abstract}

\keywords{accretion, accretion disks --- Galaxy: center --- pulsars: general --- stars: neutron --- X-rays: binaries --- X-rays: individual (XMM J174457--2850.3)}


\section{Introduction}
Low-mass X-ray binaries (LMXBs) and millisecond radio pulsars (MSRPs) are two different manifestations of neutron stars in binary systems that are thought to be evolutionarily linked \citep[e.g.,][]{alpar1982,bhattacharya1991,strohmayer1996,wijnands1998}. 

In an LMXB configuration, the neutron star accretes matter from a less massive ($\lesssim1~\Msun$) companion star that overflows its Roche lobe. Transient systems spend most of their time in a dim quiescent state with a 2--10 keV luminosity of $L_{\mathrm{X}}\simeq$$10^{31}-10^{33}~\lum$, during which the transferred matter is stored in a cold accretion disk. Once sufficient mass has been accumulated, (part of) the disk becomes hot and ionized, allowing matter to accrete onto the neutron star and giving rise to an outburst with $L_{\mathrm{X}}\simeq$$10^{36}-10^{38}~\lum$ \citep[e.g.,][]{vanparadijs1996,king1996,lasota01,coriat2012_dim}. When a certain portion of the disk mass has been accreted, the binary switches back to a quiescent state during which the disk is slowly replenished until a new outburst occurs.

These outburst/quiescence cycles eventually terminate as the binary evolves and the companion star decouples from its Roche lobe so that the mass-transfer stops \citep[e.g.,][]{tauris2012}. The rapidly rotating neutron star, spun up to millisecond periods by gaining angular momentum during the accretion phases, may now emit pulsed radio emission so that the binary is observed as a MSRP.  

Recent discoveries have opened up new opportunities to investigate the LMXB/MSRP connection. There are three MSRPs that exhibit increased X-ray states during which an accretion disk is present and radio pulsations are no longer detected: \psrm/\igrbron\ in the globular cluster M28 \citep[henceforth the M28 source; e.g.,][]{papitto2013_nature,linares2014}, \psr\ \citep[e.g.,][]{archibald2009,patruno2014}, and \xss\ \citep[e.g.,][]{demartino2013,bassa2014,roy2014}. 

The M28 source, a 3.9-ms radio pulsar, suddenly became active as an LMXB in 2013: its pulsed radio emission disappeared, while its X-ray emission increased from $L_{\mathrm{X}}\simeq10^{32}$ to $\simeq10^{37}~\lum$. The detection of a type-I X-ray burst and coherent X-ray pulsations demonstrated that the strong X-ray brightening was due to matter accreting onto the surface of the neutron star \citep[e.g.,][]{papitto2013_nature,linares2014}. After $\simeq$2 months, the accretion ceased, the X-rays faded, and the radio pulsations returned \citep[][]{papitto2013_radio}. However, instead of settling back at its quiescent level, the source jumped between $L_{\mathrm{X}}\simeq$$10^{32}$ and $10^{33}~\lum$ \citep[][]{linares2014}. 

The other two neutron stars, \psr\ and \xss\ (both spinning in 1.7 ms), display faint X-ray emission of $L_{\mathrm{X}}\simeq$$10^{32}~\lum$ and no optical signatures of an accretion disk when active as a MSRP \citep[e.g.,][]{archibald2009,bassa2014,bogdanov2011,bogdanov2014,roy2014}. However, both also exhibit an enhanced X-ray state during which no radio pulsations are detected, the X-ray luminosity is increased to $L_{\mathrm{X}}\simeq$$10^{33}~\lum$, and an accretion disk is visible in the optical band \citep[e.g.,][]{archibald2009,demartino2013,linares2014_redbacks,patruno2014}. Nevertheless, their X-ray luminosity is much lower than that of active LMXBs (i.e., $L_{\mathrm{X}}\simeq$$10^{36}-10^{38}~\lum$ such as seen for the M28 source), but rather in the regime of quiescent LMXBs. Both sources seem to spend years at a time in the two different states \citep[e.g.,][]{bassa2014,patruno2014}.

The remarkable X-ray variability displayed by these three ``transitional objects'' is thought to be related to the interaction between the neutron star magnetosphere and the accretion flow \citep[e.g.,][]{archibald2009,linares2014,papitto2014,patruno2014}. In response to fluctuations in the mass-accretion rate, the magnetospheric radius $r_m$ (the radius at which the magnetic field is dynamically important in the governing the accretion flow) may shrink or expand, giving rise to different X-ray states \citep[e.g.,][and references therein]{lipunov1992}:

\begin{enumerate}[i)]
\item For sufficiently high accretion rates, $r_m$ is smaller than the co-rotation radius $r_c$ (the radius at which the Keplerian angular velocity equals the spin velocity of the neutron star) so that accretion onto the stellar surface is centrifugally allowed. This state would correspond to the bright LMXB accretion outburst seen in the M28 source. If the matter is channeled onto the magnetic poles, pulsed X-ray emission can be observed \citep[as was indeed the case for the M28 source;][]{papitto2013}.

\item When the accretion rate drops, $r_m$ may expand beyond $r_c$ so that a propeller mechanism can operate: the fast rotating magnetic field impedes the accretion of matter onto the neutron star surface. Consequently, the X-ray luminosity is significantly lower. The interaction of the in-falling matter with the magnetosphere may suppress the radio-pulsar mechanism. This scenario could correspond to the enhanced X-ray state ($L_{\mathrm{X}}\simeq$$10^{33}~\lum$) seen in the three transitional objects. The prominent (un-pulsed) gamma-ray emission seen for \psr\ and \xss\ \citep[][]{hill2011,stappers2013} may fit into this picture, since it could result from shocks formed when the accretion stream runs into the neutron star magnetosphere \citep[e.g.,][]{papitto2014}.

\item For very low accretion rates, $r_m$ may expand beyond the light cylinder (the radius at which the co-rotation speed equals the speed of light). The pressure of the neutron star's relativistic particle wind is then sufficient to prevent matter from reaching the magnetosphere. This may clear out the accretion disk \citep[although see][]{eksi2005} and allow the radio pulsar mechanism to operate. Hence, this would correspond to the MSRP state. The faint X-ray emission ($L_{\mathrm{X}}\simeq$$10^{32}~\lum$) may find its origin in an intra-binary shock \citep[e.g.,][]{bogdanov2011,bogdanov2014,linares2014}.

\end{enumerate}

Here we discuss the peculiar X-ray properties of \xmmbron: an unclassified transient X-ray source located $\simeq$$14'$ NW of \sgra, that was discovered in 2001 \citep[][]{sakano05}. We report on the \swift\ detection of an energetic type-I X-ray burst, which unambiguously demonstrates that the source harbors an accreting neutron star, most likely in an LMXB configuration \citep[see also][]{degenaar2012_xmmburst}. By investigating 12 yr of X-ray monitoring data of the Galactic center (2000--2012), we show that it behaves similar to the M28 source.


\section{Observations and Data Analysis}
In the past decade, the Galactic center region has been targeted many times with high-resolution X-ray imaging/spectroscopic instruments onboard \swift, \chan, and \xmm. We searched the public data archives of these three observatories to characterize the long-term soft (2--10 keV) X-ray behavior of \xmmbron.  

\subsection{\swift\ XRT and BAT data}
\swift's X-Ray Telescope \citep[XRT;][]{burrows05} can acquire two-dimensional images in photon counting (PC) mode. However, for very bright sources, the instrument is operated in windowed timing (WT) mode in which the central CCD columns are collapsed into one dimension \citep[][]{hill2004}. We found 179 XRT/PC observations that had \xmmbron\ in the field of view (FOV). These $\simeq$1~ks observations were carried out between 2007 July 5 and 2012 October 31, and resulted in $\simeq181$~ks of accumulative exposure time. 

Reduction and analysis was carried out using the \swift\ tools and calibration data within \textsc{heasoft} (ver. 6.13). After processing the data with \textsc{xrtpipeline}, we employed \textsc{XSelect} to extract count rates, light curves, and spectra. We used a $15''$ circular region to obtain source events, and a $30''$ circular region for the background. The detection of a type-I X-ray burst (Section~\ref{sec:burst}) caused pile-up in the PC data. For that observation (ID 91408042) we therefore used an annular region with an inner/outer radius of $25''$/$75''$. To study the X-ray burst properties, we also analyzed the subsequent WT mode observation (ID 530588000), for which we extracted source counts from a $100''$$\times$$20''$ rectangular box. Ancillary response files were generated with \textsc{xrtmkarf} and the response matrix files (ver. 14) were sourced from the \textsc{caldb}. Using \textsc{grppha}, the spectra were grouped to a minimum of 15 photons per bin.

Due to the intense brightening caused by the type-I X-ray burst, the Burst Alert Telescope \citep[BAT;][]{barthelmy05} triggered on \xmmbron\ \citep[trigger 530588;][]{barlow2012}. We reprocessed this data and created standard data products using the \textsc{batgrbproduct} tool. The BAT is sensitive in the 15--150 keV range, but we restricted our analysis to the 15--35 keV band because the data were dominated by the background at higher energies. A spectrum was extracted with \textsc{batbinevt}, and standard corrections were applied with \textsc{batupdatephakw} and \textsc{batphasyserr}. The response matrix was created using the task \textsc{batdrmgen}.

\subsection{\xmm\ EPIC data}
We used 43 \xmm\ observations obtained with the European Photon Imaging Camera (EPIC), which consists of one PN and two MOS detectors \citep[][]{turner2001_mos,struder2001_pn}. These data span a time between 2000 September and 2012 September, with exposures ranging between $\simeq$6 and 106~ks per observation. Reduction and retrieval of data products was achieved using \textsc{SAS} (ver. 13.0). After processing the data with \textsc{epproc} and \textsc{emproc} we extracted count rates using \textsc{eregionanalyse}. To obtain source events we used a circular region with a $10''$ radius, while an aperture of twice that size was used for the background.  Spectra and response files were obtained using \textsc{specextract}. A combined spectrum of all three detectors was created using \textsc{epic$\_$spec$\_$combine}. The spectral data were grouped into bins with at least 20 photons.

\subsection{\chan\ ACIS and HRC data}
A total of 24 \chan\ observations were used, obtained with the Advanced CCD Imaging Spectrometer \citep[ACIS;][]{garmire2003_acis} or the High-Resolution Camera \citep[HRC;][]{kenter2000_hrc}. Of the two, only the ACIS provides spectral information. The \chan\ data covered the epoch between 2001 July and 2008 July, and had exposure times of $\simeq$4--167~ks per individual observation. 
Reduction and analysis were performed using the \textsc{ciao} tools (ver. 4.5). After initial reprocessing, we used \textsc{dmextract} to obtain source counts from a circular region with a radius of $5''$, and background counts from a region with a radius of $10''$. We extracted source and background spectra from the ACIS data using \textsc{specextract}, which also generated the response files. The spectra were grouped to a minimum of 20 photons~bin$^{-1}$.

\subsection{Spectral Analysis and Light Curve Construction}
The spectral data were modeled using \textsc{XSpec} \citep[ver. 12.8;][]{xspec}. Fits were restricted to the 2--10 keV range (15--35 keV for the BAT) because of the very high hydrogen column density ($N_{\mathrm{H}}\simeq10^{23}~\nh$). We used a black body model (\textsc{bbodyrad}) to describe the type-I X-ray burst (Section~\ref{sec:burst}), whereas a power-law model (\textsc{pegpwrlw}, with the normalization set to represent the 2--10 keV range) was used for all other data. The \textsc{tbabs} model was included to account for interstellar absorption, using the \textsc{vern} cross-sections and \textsc{wilm} abundances \citep[][]{verner1996,wilms2000}. 

We deduced unabsorbed fluxes in the 2--10 keV energy range and converted these into luminosities by assuming a distance of $D=6.5$~kpc (see Section~\ref{subsec:burstspec}). To obtain 2--10 keV flux estimates for observations with a low number of counts and to obtain flux upper limits for non-detections, we determined instrument-specific count rate to flux conversion factors. Throughout this work, quoted fluxes and luminosities refer to the 2--10 keV band unless stated otherwise. Errors reflect 90\% confidence levels.

\begin{table}[]
\begin{center}
\caption{Observations for X- Ray Spectral Analysis.\label{tab:spec}}
\begin{tabular*}{0.50\textwidth}{@{\extracolsep{\fill}}cccc}
\hline\hline
Epoch  & Instr. & ObsID & State  \\
 &  &  &  \\
\hline
2002 May & \chan & 3392/93 & Q  \\
2004 Aug/Sep & {\it XMM} & 0202670701/801 & Q   \\
2007 Mar/Apr & {\it XMM} & 0402430301/401/701 & I  \\
2011 Mar/Apr & {\it XMM} & 0604300601/701/801/901/1001 & I   \\
2012 Mar & {\it XMM} & 0674600601/701/801/1001/1101 & I   \\
2001 Sep & {\it XMM} & 0112972101 & O   \\
2008 Jun & \swift & 0035650109/110 & O   \\
2012 Aug & \swift & 0091408042/43/44 & O   \\
\hline
\end{tabular*}
\tablecomments{The different 2--10 keV luminosity states are indicated as Q (quiescence; $L_{\mathrm{X}}\lesssim10^{33}~\lum$), I (intermediate; $L_{\mathrm{X}}\simeq10^{33}-10^{34}~\lum$), and O (outburst; $L_{\mathrm{X}}\simeq10^{34}-10^{36}~\lum$). 
}
\end{center}
\end{table}

\section{Accretion History}\label{subsec:acchist}

 \begin{figure*}
 \begin{center}
 	\includegraphics[width=10cm]{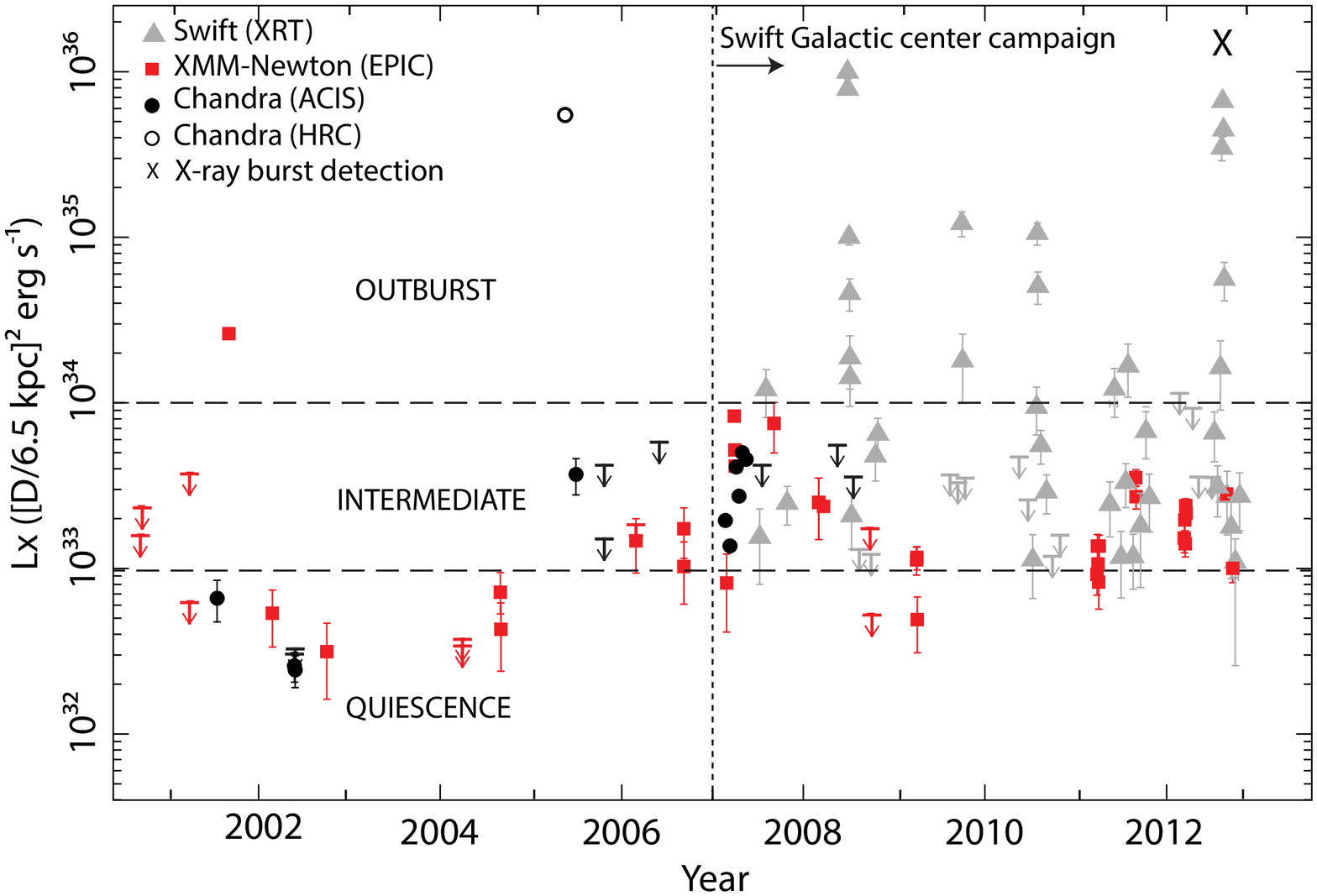}\hspace{+1.0cm}
 	\includegraphics[width=7cm]{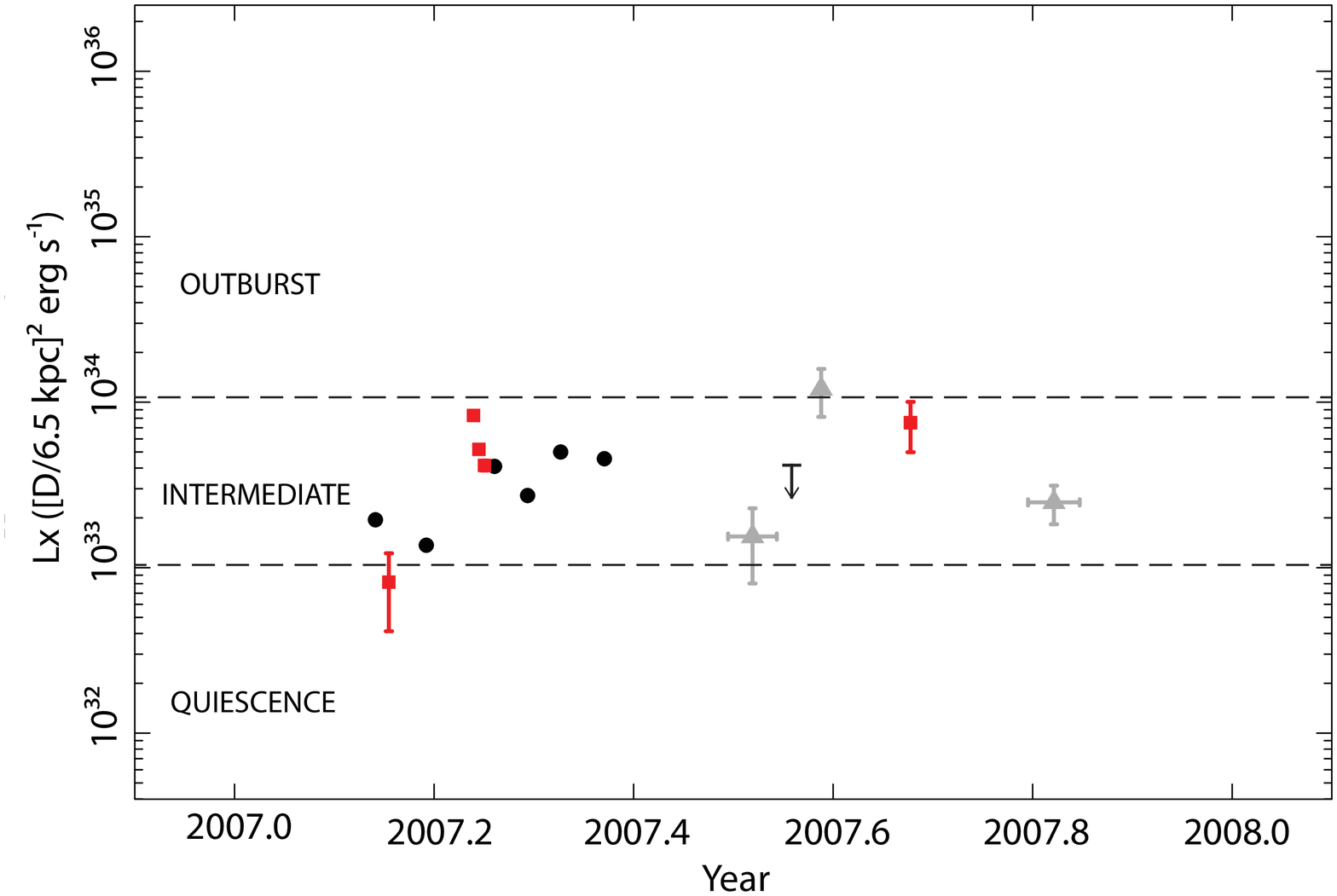}\hspace{+1.0cm}
 	\includegraphics[width=7cm]{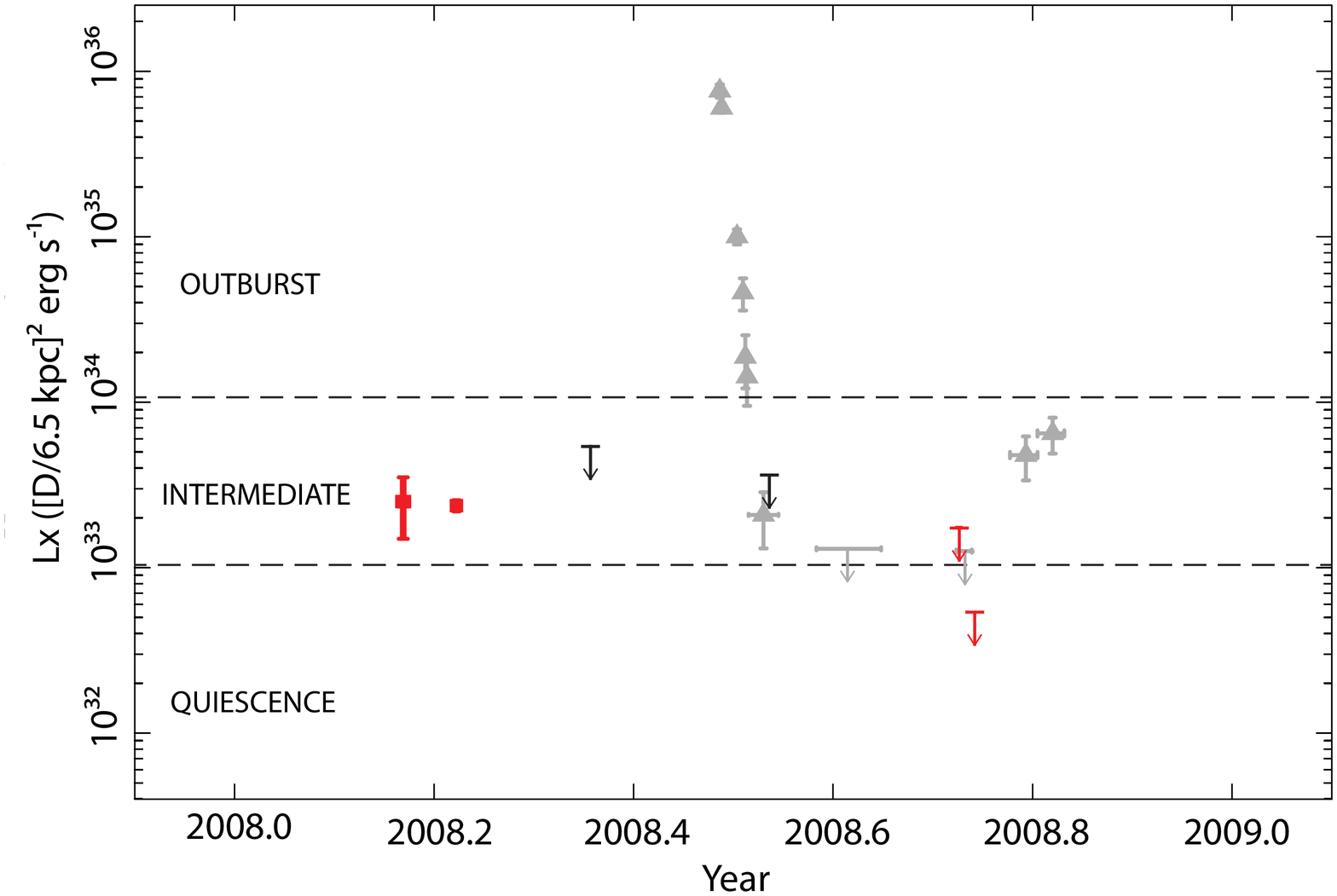}
 	\includegraphics[width=7cm]{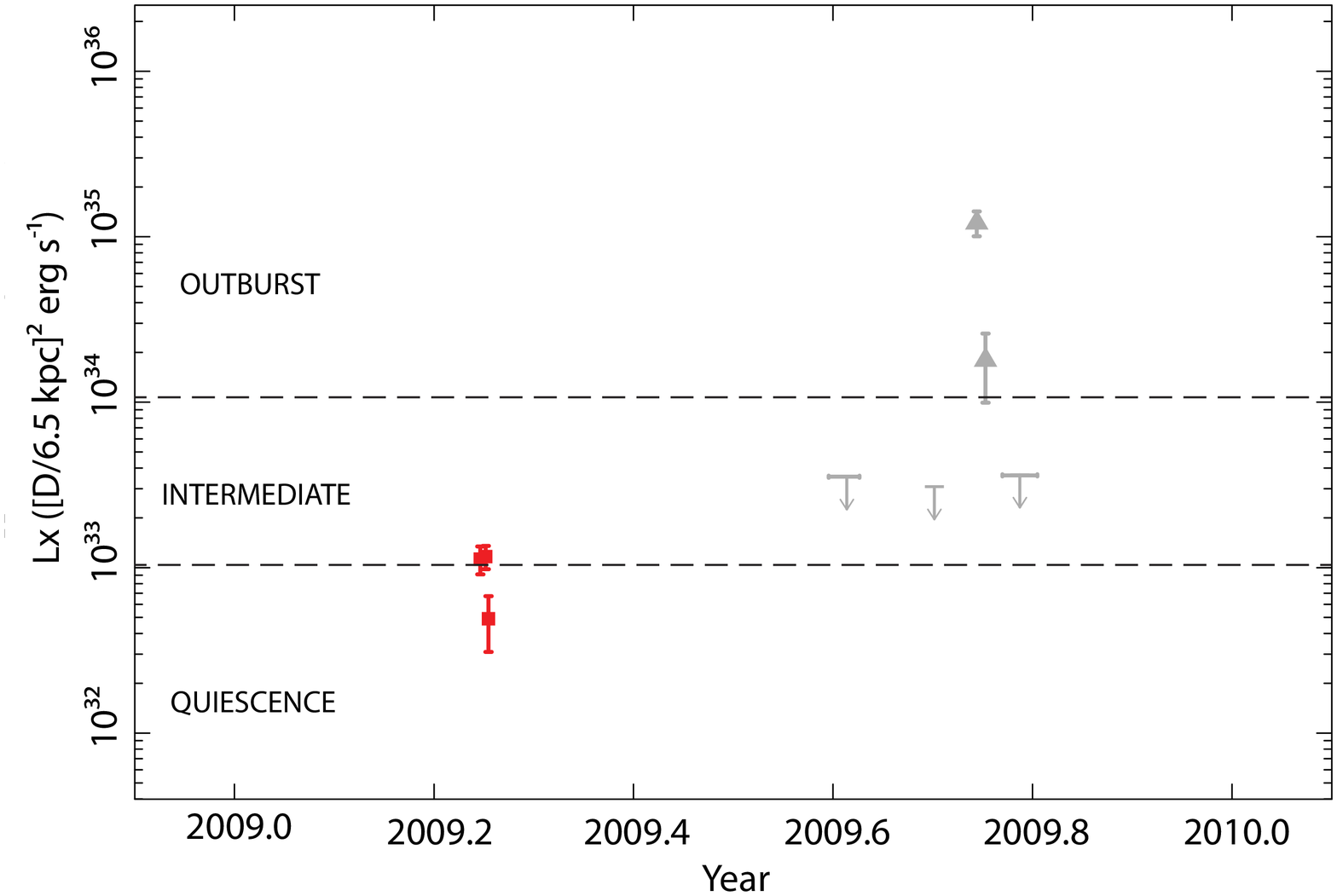}\hspace{+1.0cm}
 	\includegraphics[width=7cm]{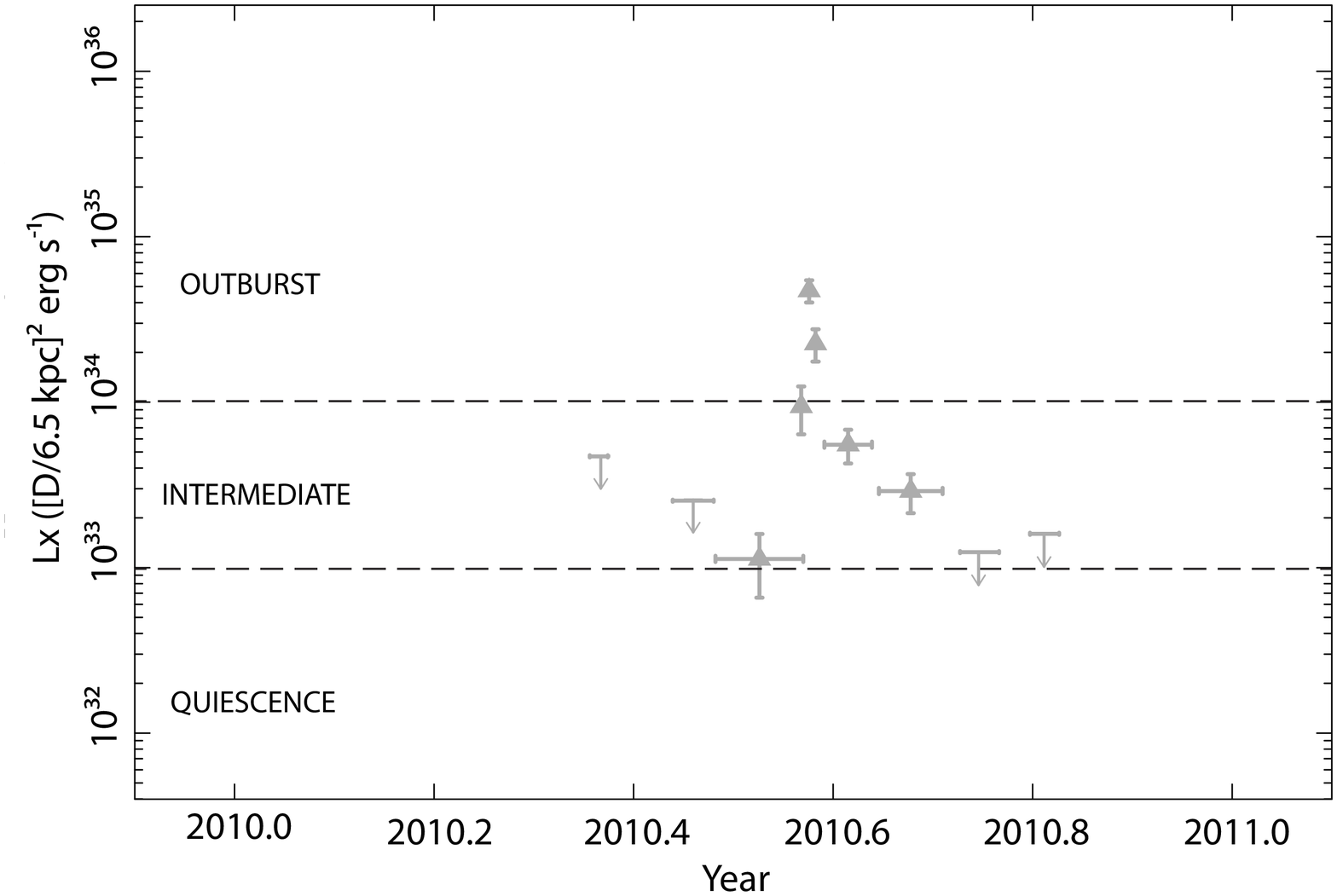}
 	\includegraphics[width=7cm]{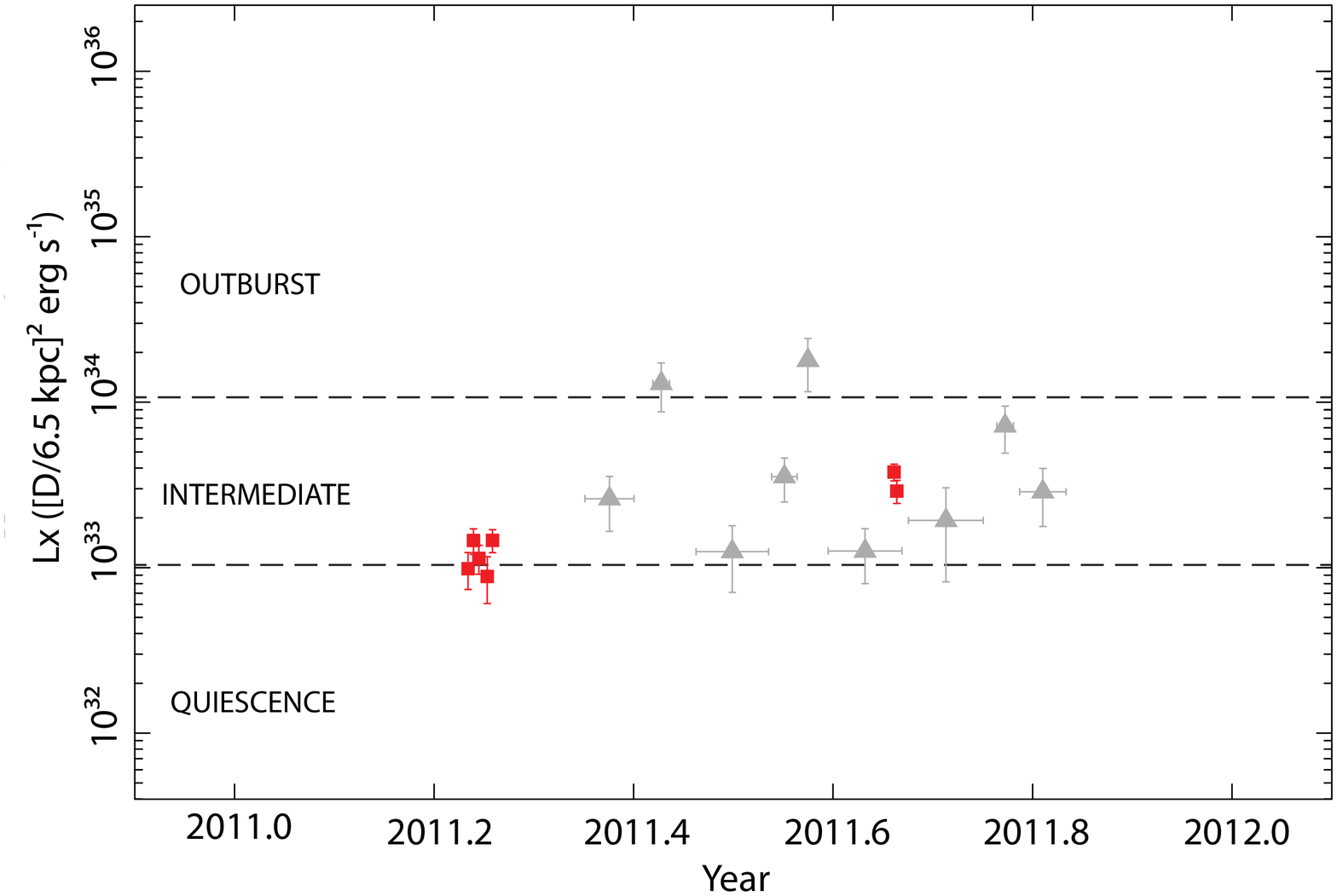}\hspace{+1.0cm}
 	\includegraphics[width=7cm]{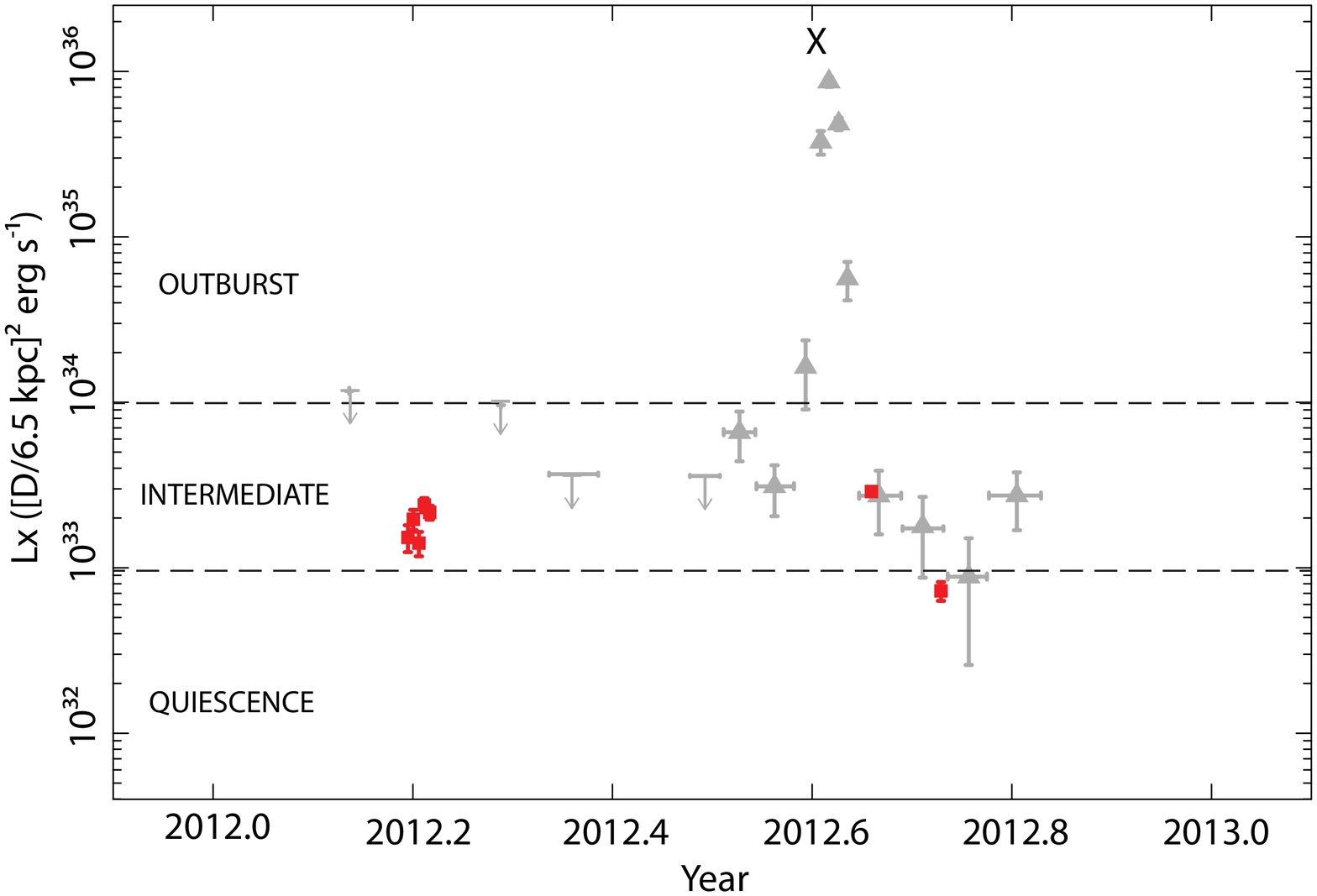}	
    \end{center}
    \caption[]{X-ray light curves (2--10 keV). The top panel displays the long-term (2002--2012) luminosity evolution, whereas the six lower panels provide expanded views of the 2007--2012 epoch. \swift\ data is indicated by grey triangles, \xmm\ by red squares, and \chan\ by black circles (where open and filled circles correspond to HRC and ACIS data, respectively). Upper limits represent 2$\sigma$ confidence levels, error bars reflect 90\% confidence intervals. The ``X'' symbol marks the time of the X-ray burst detection with \swift. The horizontal dashed lines provide a handle on the different luminosity states exhibited by the source. The vertical dotted line in the top panel indicates intensified monitoring of the source through the \swift\ Galactic center campaign.}
 \label{fig:longlc}
\end{figure*}

\subsection{Long-term X-Ray Light Curve}\label{subsec:longlc}
To investigate the long-term accretion history of \xmmbron, we assembled a total of 246 \swift, \xmm, and \chan\ observations obtained between 2000 September and 2012 October. To increase the sensitivity of the short \swift\ exposures, we added chunks of data performed within a time frame of $\simeq$2--4~weeks that added up to $\simeq$5--10~ks of exposure time (except during outbursts when the source was readily detected within single observations). 

Figure~\ref{fig:longlc} (top) reveals at least five distinct outbursts reaching $L_{\mathrm{X}} \simeq 10^{35}-10^{36}~\dist~\lum$ over the past 12 yr: in 2005, 2008, 2009, 2010, and 2012 \citep[][]{wijnands06,degenaar2010_gc,degenaar2010_atel_grs_xmm,degenaar2012_gc,degenaar2012_xmmactive}. The type-I X-ray burst was detected during the 2012 outburst (Section~\ref{sec:burst}). Owing to the high cadence of observations, four of the outbursts were caught by the \swift\ monitoring program. This provides constraints on the duration of the outbursts: $\simeq$1--7 weeks in 2008, $\simeq$1--9 days in 2009, $\simeq$5--21~days in 2010, and $\simeq$15--21~days in 2012 \citep[see also][]{degenaar2010_gc,degenaar2012_gc}. 

The lowest luminosity exhibited by the source, as seen with \chan\ and \xmm\ in 2002 and 2004, is $L_{\mathrm{X}} \simeq (2.1-5.5) \times 10^{32}~\dist~\lum$ \citep[see also][]{sakano05}. However, it is not often found at its quiescent level, but is instead frequently detected at  $L_{\mathrm{X}} \simeq 10^{33}-10^{34}~\dist~\lum$, i.e., intermediate between quiescence and full outburst. 
This is perhaps best illustrated by the expanded view of the 2007, 2011, and 2012 light curves, when the source was most densely sampled (Figure~\ref{fig:longlc}). 

In 2007 February--May, the source was covered during ten \chan/\xmm\ observations, which showed that its luminosity varied between $L_{\mathrm{X}}\simeq(0.8-8)\times10^{33}~\dist~\lum$. These observations were spaced by 2--18 days, making it unlikely that all these caught the rise/decay of bright outbursts that were missed. Rather, it seems to suggests that the source lingered in an intermediate luminosity state for several months. This is supported by dense monitoring in 2011 May--November (on average two observations per week, with gaps of only 2--10 days), which showed the source hovering at $L_{\mathrm{X}}\simeq(1-20)\times10^{33}~\dist~\lum$ for several months while it did not reach quiescent or bright outburst levels during that time (Figure~\ref{fig:longlc}). It is also striking that in 2012 the source was already detected well above quiescence at $L_{\mathrm{X}} \simeq 5 \times 10^{33}~\dist~\lum$ for $\simeq$1 month prior to its bright outburst, and instead of decaying  to quiescence afterward, it remained at $L_{\mathrm{X}} \simeq 10^{33}~\dist~\lum$ for at least another 2 months.

Examination of the long-term light curve thus shows that the low-activity stages can last for several months, and that these are not necessarily associated with bright outbursts \citep[see also][]{degenaar2010_gc,degenaar2012_gc}. 
It therefore appears that the source occupies three distinct luminosity regimes: (1) accretion outbursts of $L_{\mathrm{X}} \simeq10^{34}-10^{36}~\dist~\lum$, (2) a quiescent state with $L_{\mathrm{X}}\lesssim10^{33}~\dist~\lum$, and (3) an intermediate state characterized by $L_{\mathrm{X}}\simeq10^{33}-10^{34}~\dist~\lum$ (2--10 keV). 


We note that the number of source counts detected in individual observations is low: $\simeq$$0.002-0.3~\cnts$ for \xmm, $\simeq$$0.0005-0.01$ (ACIS) and $\simeq0.3~\cnts$ (HRC) for \chan, and $\simeq$$0.001-0.3~\cnts$ for \swift. Moreover, the source was often found at the edge of the detector, which significantly decreases the effective exposure time. The low statistics (and the low timing resolution of the imaging observations) did not allow us to conduct meaningful searches for (fast) X-ray pulsations.\footnote[9]{\citet{sakano05} reported the possible detection of slow ($\simeq$5.2~s) X-ray pulsations during a 2001 \xmm\ observation in which the source was detected at $L_X\simeq2\times10^{34}~\lum$ (ObsID 0112972101). However, there are no details given on the search methods or the significance of the signal. We were not able to reproduce these results and did not obtain any meaningful limits.}

 \begin{figure*}
 \begin{center}
 	\includegraphics[width=18cm]{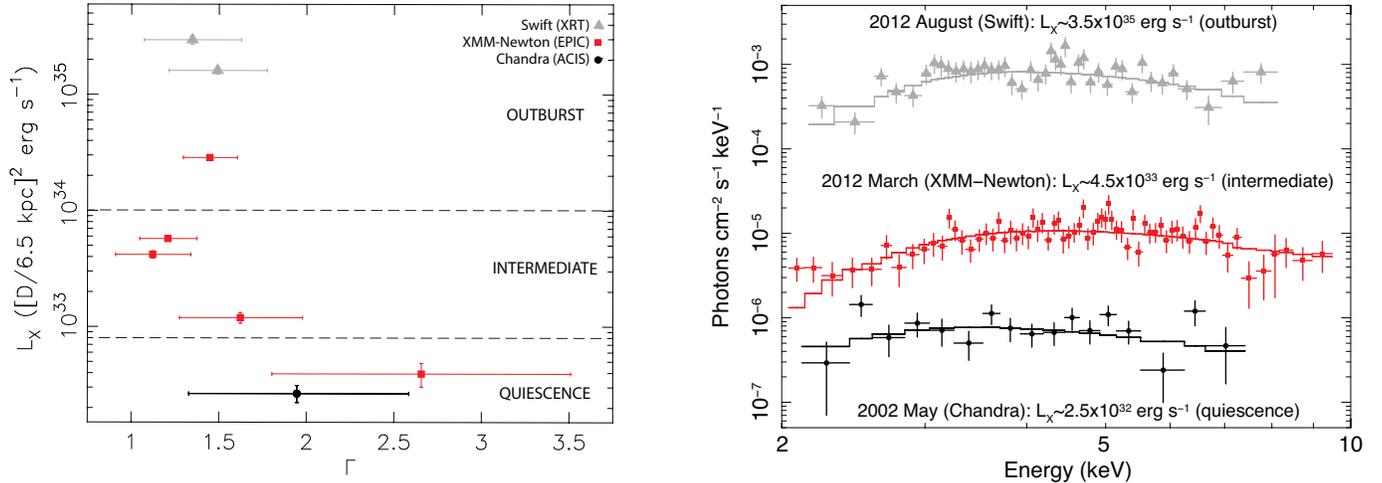}
    \end{center}
    \caption[]{Spectral evolution. \chan\ data is indicated by black filled circles, \xmm\ data by red squares, and \swift\ data by grey triangles. Error bars represent 90\% confidence intervals. Left: photon index ($\Gamma$) vs. 2--10 keV luminosity ($L_{\mathrm{X}}$). Right: a representative selection of unfolded X-ray spectra of different luminosity states. The solid lines indicates fits to an absorbed power-law model. }
 \label{fig:obspec}
\end{figure*}

\subsection{X-Ray Spectral Properties}\label{subsec:specob}
\xmmbron\ is so faint that most observations did not collect sufficient photons to allow for spectral analysis. Nevertheless, we obtained a few good spectra to characterize the different luminosity regimes and to investigate any possible changes in spectral shape (Table~\ref{tab:spec}). To improve the statistical quality of the spectra we summed several observations (of the same instrument) that had similar source count rates.

For the outburst state we use three epochs: \xmm\ data obtained in 2001, and \swift\ data covering the 2008 and 2012 outbursts.\footnote[10]{The 2005 outburst was captured by the \chan/HRC and hence no spectral information is available. The 2009 and 2010 \swift\ outbursts were only sparsely sampled and did not yield as good quality spectra as the 2008 and 2012 outbursts (Figure~\ref{fig:longlc}).} Two different epochs were used to characterize the quiescent state: we summed the spectra of two long ($\simeq$157--167~ks) \chan/ACIS observations performed on 2002 May 25/28, and combined two long ($\simeq$135~ks) \xmm\ exposures of 2004 August 31 and September 2. Finally, the intermediate luminosity state is best covered by a series of \xmm\ exposures obtained during three different epochs: 2007 March 30 till April 3, 2011 March 28 till April 5, and 2012 March 13--31 (with exposure times of $\simeq$16--106~ks). 

This set of eight spectra were fitted together to an absorbed power-law model. We first assumed that the hydrogen column density and photon index remained the same at all epochs, allowing only the normalization of the \textsc{pegpwrlw} model (the unabsorbed 2--10 keV flux) to vary between the different observations. This resulted in a reasonable fit ($\chi^{2}_{\nu}=1.11$ for 554 dof) with $N_{\mathrm{H}}=(1.1\pm 0.1)\times10^{23}~\nh$, and $\Gamma=1.43 \pm 0.14$. Allowing either $N_{\mathrm{H}}$ or $\Gamma$ to vary between different epochs improved the fit ($\chi^{2}_{\nu}=1.07/1.08$ for 554 dof), although an f-test suggests that there is a $0.0003/0.003$ probability that such an improvement is achieved by chance. Indeed, there appears to be little variation in $\Gamma$ and $N_{\mathrm{H}}$ between different epochs. This is illustrated by Figure~\ref{fig:obspec} (left), where we show the values of $\Gamma$ obtained for the fit in which $N_{\mathrm{H}}$ was tied between the different data sets (yielding $N_{\mathrm{H}}=(1.1\pm0.1)\times10^{23}~\nh$). The spectrum seems to be somewhat softer in quiescence than at higher luminosities, although the 90\% error bars are large and partly overlap with those of the brighter states. 

It seems that despite four orders of magnitude change in luminosity, \xmmbron\ shows no strong spectral variability in the 2--10 keV energy range. This can also be seen in Figure~\ref{fig:obspec} (right), which displays three spectra that are representative of the different luminosity states. We note that due to the large absorption column in the direction of the source, we cannot obtain any meaningful upper limits on the quiescent thermal emission from the surface of the neutron star.

\section{Detection of a Type-I X-ray Burst}\label{sec:burst}
Regular monitoring of the Galactic center with \swift/XRT showed that \xmmbron\ started a new accretion outburst around 2012 August 5 \citep[Figure~\ref{fig:longlc};][]{degenaar2012_xmmactive}. Six days later, on August 11, the BAT triggered on \xmmbron\ \citep[trigger 530588;][]{barlow2012}. Preliminary analysis suggested that this trigger was caused by a type-I X-ray burst \citep[][]{degenaar2012_xmmburst}: a bright flash of X-ray emission resulting from unstable thermonuclear burning of He and/or H on the surface of a neutron star. These explosive phenomena are characterized by a sharp rise in X-ray luminosity (up to the Eddington limit) resulting from  rapid ignition of the fuel layer, followed by a slower decay tail as the burning ashes cool. 

Below, we present the detailed analysis of the BAT trigger on \xmmbron. The light curve morphology (a double peak followed by a long decay along a power law with an index of $-\alpha$$\simeq$1.05), the spectral shape and evolution (a black body cooling from $kT$$\simeq$2 to $\simeq$1~keV), and the estimated bolometric energy output ($E_b\simeq5\times10^{40}$~erg) are indeed all consistent with a type-I X-ray burst from an accreting neutron star LMXB. 

\subsection{X-Ray Burst Light Curve Analysis}\label{subsubsec:lcana}
The 15--35 keV BAT light curve of trigger 530588, displayed in Figure~\ref{fig:lc}, shows an enhancement above the background for $\simeq$75~s. The main peak lasts $\simeq$20~s and is centered on the trigger time ($t_{0}=0$; 2012 August 11 at 04:44 \textsc{utc}), but there is also a much sharper peak $\simeq$50~s before the BAT trigger (see the inset in Figure~\ref{fig:lc}). A double-peaked structure is often observed for type-I X-ray bursts and could be a signature of a photospheric radius expansion phase that occurs when the generated emission reaches the Eddington limit.

By coincidence, the BAT trigger occurred when \swift/XRT was observing the Galactic center as part of its regular monitoring program (ObsID 91408042). The $\simeq$85-s long PC mode observation shows that \xmmbron\ was the only active X-ray source in the FOV. Strikingly, the source was a factor $\simeq$1000 brighter than during the preceding $\simeq$300-s exposure that was obtained $\simeq$3.5~hr earlier (when it was detected at $\simeq0.3~\cnts$). This indicates that the peak of the X-ray burst was also caught by the XRT (see Figure~\ref{fig:lc}).

In response to the BAT trigger, the PC mode observation was interrupted, the spacecraft slewed, and observations of \xmmbron\ resumed with the XRT operating in WT mode $\simeq$75~s later. The $\simeq510$~s exposure showed that the source intensity steadily decayed from $\simeq$30 to $\simeq 4~\cnts$ (Figure~\ref{fig:lc}). There is no indication that the decay was leveling off at the end of the XRT observation, and the source was still a factor $\simeq$10 brighter than observed $\simeq$3.5~hr prior to the BAT trigger ($\simeq0.3~\cnts$). This suggests that the cooling tail of the X-ray burst was ongoing and thus had a length of $>$10~minutes.

To characterize the decay of the X-ray burst we fitted the WT light curve to a power law of the form $y=A\times(t-t_{0})^{-\alpha}$. This yielded an index of $-\alpha=1.05\pm 0.02$ and normalization of $A=3371\pm 390~\cnts$ ($\chi^2_{\nu}=1.51$ for 100 dof; dotted curve in Figure~\ref{fig:lc}). Extrapolating the power-law decay down to the pre-burst level of $\simeq0.3~\cnts$, we estimate that the X-ray burst may have been as long as $\simeq$7400~s ($\simeq$2.1~hr). 
We searched the 0.5--10 keV XRT/WT for burst oscillations \citep[modulations at the neutron star spin period; e.g.,][]{strohmayer06} but found none, with a $3\sigma$ upper limit of 6\%. 

\subsection{X-Ray Burst Spectral Analysis}\label{subsec:burstspec}
\begin{figure}
 \begin{center}
	\includegraphics[width=8.5cm]{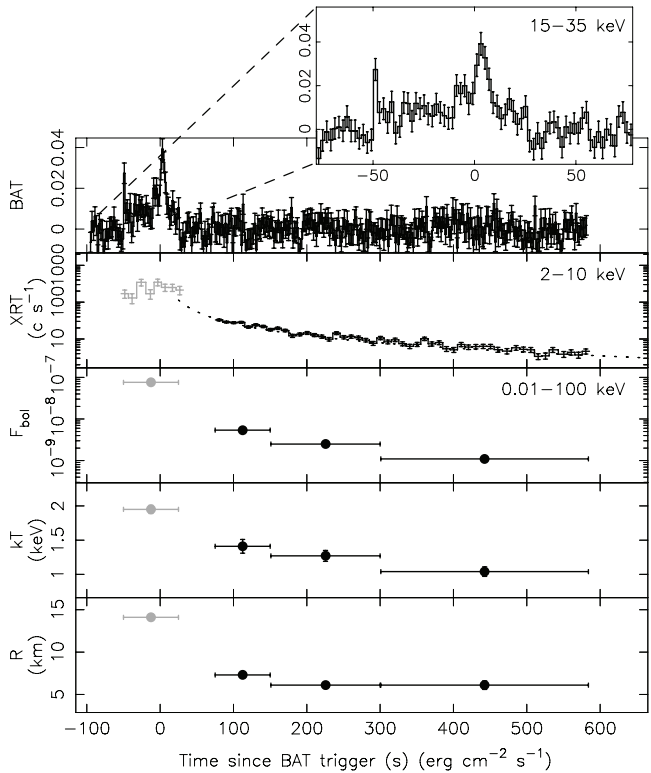}
    \end{center}
    \caption[]{Results from X-ray burst analysis. Error bars represent 90\% confidence intervals. From top to bottom: BAT 15--35 keV light curve of trigger 530588 at 2~s resolution, XRT 2--10 keV light curve at 10~s resolution consisting of PC (gray) and WT (black) data, 0.01--100 keV unabsorbed flux, black body temperature, and emitting radius (for $D=6.5$~kpc). The dotted line in the second panel represents a power-law decay fit to the X-ray burst tail. }
 \label{fig:lc}
\end{figure}

We extracted a BAT spectrum using 75 s of data obtained from $t$=$-50$~s to $t=+25$~s since the trigger time. The XRT/PC monitoring observation ran from $t=-57$~s to $t=+28$~s, hence coinciding with the BAT detection (see Figure~\ref{fig:lc}). To exploit this broad-band coverage, we fitted the XRT/PC and BAT spectra simultaneously (each with their own response files), treating them as a single spectrum. The follow-up XRT/WT data were divided into three intervals of $\simeq$1500 counts each.

We fitted the X-ray burst spectra together to an absorbed black body with the hydrogen column density ($N_{\mathrm{H}}$) tied between the different intervals. To account for the underlying accretion emission we included a power-law component with the parameters fixed to the values obtained from fitting the spectrum of the $\simeq$300-s PC mode data that preceded the X-ray burst detection (a photon index of $\Gamma = 2.0$ and a 2--10 keV unabsorbed flux of $F_{\mathrm{X}} = 8.5\times10^{-11}~\flux$). 

This approach resulted in a good fit ($\chi^2_{\nu}=0.99$ for 333 dof). The joint value of $N_{\mathrm{H}}=(9.2\pm0.7)\times10^{22}~\nh$ is consistent with that found for accretion outbursts of the source (Section~\ref{subsec:specob}). For the peak of the X-ray burst, we infer a temperature of $kT=1.95\pm0.06$~keV, which decays to $kT=1.04\pm0.07$~keV in the final part of the XRT/WT observation (Figure~\ref{fig:lc}). Extrapolating the black body fits to an energy range of 0.01--100 keV we estimate a bolometric peak flux of $F_{\mathrm{bol}}\simeq(7.8\pm 0.5)\times10^{-8}~\flux$. 

The duration and energy output suggests that we observed a pure He X-ray burst (see Section~\ref{subsec:peculiar}), which are thought to reach the Eddington luminosity ($L_{\mathrm{Edd}}$). The double-peaked structure seen in the BAT data may support this (Section~\ref{subsubsec:lcana}). Therefore, assuming that the bolometric peak flux corresponded to the empirical value of $L_{\mathrm{Edd}}=3.8\times10^{38}~\lum$ \citep[][]{kuulkers2003}, we estimate a source distance of $D=6.5\pm 0.2$~kpc.

\subsection{X-Ray Burst Energetics}\label{sec:energy}
For $F_{\mathrm{bol}}=(7.8\pm 0.5)\times10^{-8}~\flux$, the fluence in the $\simeq$75 s X-ray burst peak is $f_{\mathrm{peak}}\simeq5.7\times10^{-6}~\fluence$. The energy emitted during the X-ray burst tail can be estimated by integrating the power-law decay fit from $t = +25$~s (the time when the peak had decayed back into the BAT background; Figure~\ref{fig:lc}) to $t = +7400$~s after the BAT trigger (the time at which the XRT count rate had presumably decayed to the persistent level; Section~\ref{subsubsec:lcana}). This yields $f_{\mathrm{tail}}\simeq3.8\times10^{-6}~\fluence$, and hence a total fluence for the X-ray burst of $f_{\mathrm{b}}\simeq9.5\times10^{-6}~\fluence$. For $D=6.5$~kpc, this implies a total radiated energy output of $E_{\mathrm{b}}\simeq4.7\times10^{40}$~erg. Table~\ref{tab:burst} summarizes the X-ray burst properties.

The nuclear energy generation rate for a pure He burst is $Q_{\mathrm{nuc}} \simeq 1.6$~MeV~nucleon$^{-1}\simeq1.5\times10^{18}$~erg~g$^{-1}$ \citep[e.g.,][]{galloway06}. This suggests that the ignition layer had a mass of $E_{\mathrm{b}}(1+z)/Q_{\mathrm{nuc}}\simeq4\times10^{22}$~g (where $1+z$=1.31 is the gravitational redshift for a neutron star with $R=10$~km and $M=1.4~\Msun$). From the flux measured $\simeq$3.5~hr prior to the X-ray burst we estimate that the source was accreting at a bolometric luminosity of $L_{\mathrm{acc}} \simeq 1.3 \times10^{36}~\dist~\lum$, i.e., $\simeq$0.3\% of the Eddington limit \citep[where we assumed a bolometric correction factor of $\simeq$3 for converting from the 2--10 keV band flux;][]{zand07}. This corresponds to a mass-accretion rate of $\dot{M}=RL_{\mathrm{acc}}/GM \simeq 7\times10^{15}~\mdotgs$ (where $G$ is the gravitational constant). Powering the observed X-ray burst would thus require the source to accrete at this level for $\simeq$10~weeks. Since this is longer than the observed duration of the bright accretion episodes (see Section~\ref{subsec:longlc}), it would take several outbursts (i.e., several years) to build up enough fuel.


\section{Discussion}

\subsection{An LMXB with Peculiar X-Ray Behavior}\label{subsec:peculiar}
\xmmbron\ was tentatively identified as a neutron star LMXB based on the amplitude of its X-ray variability and the lack of a bright radio counterpart \citep[][]{sakano05,wijnands06}. In this work, we have reported the detection of a type-I X-ray burst, which confirms this classification. Based on the observed peak flux we estimate a source distance of $D=6.5$~kpc. 

The observable properties of type-I X-ray bursts (e.g., the duration, peak brightness and energy output) depend on the conditions of the ignited fuel layer (e.g., its H content and thickness), and are set by the accretion rate onto the neutron star \citep[][]{fujimoto81,bildsten98,peng2007}. The long cooling tail ($>$10~minutes, up to $\simeq2.1$~hr) and high radiated energy output ($E_b\simeq5\times10^{40}$~erg) observed for \xmmbron\ fall into the regime of  intermediately long X-ray bursts. These rare events are thought to arise from ignition of a thick, pure He layer and are typically observed from neutron stars accreting slowly at $\lesssim$1\% of the Eddington limit, like \xmmbron\ \citep[e.g.,][]{zand05,falanga08,linares09,degenaar2010_burst,degenaar2011_burst,degenaar2013_igrj1706}. It takes a long time to accrete enough fuel to power these energetic X-ray bursts (several weeks/months), so it is not surprising that this was the first X-ray burst ever detected from the source, despite its frequent activity over the past decade. 

Investigation of the long-term (2000--2012) light curve revealed several accretion outbursts reaching up to $L_{\mathrm{X}}\simeq10^{35}-10^{36}~\dist~\lum$ and lasting for a few days--weeks. The brightness, duration, and repetition rate of these outbursts is not unusual for Galactic LMXBs \citep[e.g.,][]{chen97,wijnands06,degenaar2010_gc}. Its quiescent luminosity of $L_{\mathrm{X}} \simeq(2-5)\times 10^{32}~\dist~\lum$ is typical for neutron LMXBs. However, it is striking that the source is often detected at an intermediate luminosity of $L_{\mathrm{X}} \simeq 10^{33}-10^{34}~\dist~\lum$, sometimes for months at a time. Below we discuss different scenarios that may account for this peculiar X-ray variability. We consider the most likely explanation that it is caused by the interaction of the magnetic field of the neutron star with the accretion disk, similar to what has been proposed for the MSRP/LMXB transitional objects.

\subsection{Quasi-stable Accretion Disk or Wind Accretion?}
It is not immediately clear how quasi-stable low-luminosity states such as seen in \xmmbron\ can be explained within the disk instability model that ascribes outburst and quiescent cycles to changes in the ionization state of the accretion disk. There are some parallels with Z-Cam stars, a sub-class of accreting white dwarfs that undergo outburst/quiescent cycles but sometimes get caught in a ``standstill''. These states may last several days to years and are $\simeq$1~optical magnitudes fainter than the maximum intensity reached during outbursts. It is thought that the mass-accretion rate is fluctuating around a critical rate (perhaps due to sunspot activity on the donor star), so that the accretion disk can be found both in a stable (standstill) and unstable (outburst-quiescence cycles) configuration \citep[e.g.,][]{meyer1983,lasota01}.

The detection of a bright outburst from \xmmbron\ in 2012, while it was in an intermediate-luminosity state right before and after, does not fit into this picture. During a standstill, the disk should be in a stable, ionized configuration and therefore the thermal-viscous instability that causes bright accretion outbursts cannot occur. Interestingly, three Z-Cam stars were recently found to display outbursts during their standstills \citep[Iw And, V513 Cas, and ST Cha;][]{simonsen2011,simonsen2014,szkody2013}. 

This apparent challenge to the disk instability model was addressed by \citet{hameury2014}, who concluded that no physically motivated modifications to the model could account for the observations. Therefore, these authors proposed that the outbursts observed during standstills should be caused by brief enhancements in the mass-transfer rate (e.g., due to magnetic activity) of the donor star. One requirement is then that an outburst is followed by a brief but significant luminosity dip before it restores to the standstill level. This was not observed in \xmmbron, which  settled right back at its intermediate-luminosity state after its 2012 outburst. Moreover, it is not clear if accretion disks in neutron star LMXBs would respond the same as these dwarf novae.

Given the difficulty of explaining the peculiar X-ray flux behavior of \xmmbron\ in the disk instability model, it was previously proposed that it might be a wind-accreting system \citep[][]{degenaar2010_gc}. Indeed, there is a striking resemblance with Super-giant Fast X-ray Transients (SFXTs), in which a neutron star is accompanied by an O/B supergiant \citep[e.g.,][for a review]{sidoli2013}. A common property of these objects is that their outburst ($L_{\mathrm{X}}\simeq10^{36}-10^{37}~\lum$) and quiescent states ($L_{\mathrm{X}}\simeq10^{32}~\lum$) are relatively rare; they are most often detected at intermediate levels of $L_{\mathrm{X}}\simeq10^{33}-10^{34}~\lum$ \citep[e.g.,][]{romano2011}. Their behavior is explained in terms of quasi-spherical accretion from the (clumpy) winds of their companion \citep[e.g.,][]{drave2014}. A similar phenomenology is observed for Symbiotic X-ray binaries (SyXRBs), in which a neutron star accretes from the wind of an M-giant \citep[e.g.,][]{masetti2007,patel2007,enoto2014}. 

Despite the similarities in X-ray behavior, the lack of an infra-red counterpart with $K_s \lesssim15.6$~mag \citep[][]{mauerhan2009}, renders it unlikely that \xmmbron\ is a SFXT or a SyXRB \citep[it also rules out a main sequence companion with a spectral type earlier than B3;][]{degenaar2010_gc}. Moreover, the neutron stars in these systems are believed to have very strong magnetic fields ($B$$\gtrsim10^{12}$~G), which prevents the occurrence of type-I X-ray bursts such as seen for \xmmbron\ \citep[e.g.,][]{bildsten98}. It has also been proposed that LMXBs with low accretion luminosities such as \xmmbron\ are capturing the wind of an M-dwarf companion \citep[][]{maccarone2013}. However, its time-averaged mass-accretion rate of $\dot{M}$$\simeq$$10^{-12}-10^{-11}~\mdot$ \citep[][]{degenaar2010_gc} seems to be too high. Wind accretion therefore likely cannot account for the peculiar X-ray flux behavior of \xmmbron.

\begin{table}
\begin{center}
\caption{X-Ray Burst Properties.\label{tab:burst}}
\begin{tabular*}{0.49\textwidth}{@{\extracolsep{\fill}}lc}
\hline\hline
Parameter (unit)  & Value \\
\hline
Powerlaw decay index, $\alpha$ \dotfill & $-1.05\pm 0.02$  \\
Total duration, $t_{\mathrm{b}}$ (hr) \dotfill & $\simeq 2.1$  \\
Total fluence, $f_{\mathrm{tot}}$ ($\fluence$) \dotfill & $\simeq 9.5 \times 10^{-6}$   \\
Radiated energy, $E_{\mathrm{b}}$ (erg) \dotfill & $\simeq 4.7 \times 10^{40}$   \\
Bolometric peak flux, $F_{\mathrm{bol}}$ ($\flux$) \dotfill & $(7.8\pm0.5) \times 10^{-8}$   \\
Distance, $D$ (kpc) \dotfill & $6.5\pm0.2$  \\
Pre-burst accretion luminosity, $L_{\mathrm{acc}}$ ($\lum$) \dotfill & $\simeq1.3 \times 10^{36}$ \\
Pre-burst mass-accretion rate, $\dot{M}$ ($\Msun~\mathrm{yr}^{-1}$) \dotfill & $\simeq1.1 \times 10^{-10}$ \\
\hline
\end{tabular*}
\tablecomments{The bolometric peak flux and accretion luminosity are unabsorbed and for the 0.01--100 keV energy range. 
}
\end{center}
\end{table}

\subsection{Interaction with the Neutron Star Magnetic Field?}
The properties of \xmmbron\ are reminiscent of the M28 source, and the two other MSRPs that show enhanced X-ray states. The state changes and X-ray flux variations observed in these three objects are thought to be related to the interplay between the neutron star's magnetic field and the accretion disk \citep[Section~1; e.g.,][]{archibald2009,linares2014,papitto2014,patruno2014}. 

A similar mechanism might be at work in \xmmbron. If true, the source likely has a sizable magnetic field and may exhibit (fast) X-ray pulsations during accretion outbursts. However, the faintness of the source in the present data did not allow us to conduct meaningful pulsation searches. It might also exhibit radio pulsations during its quiescent X-ray state ($L_{\mathrm{X}}\simeq5\times10^{32}~\lum$). Sadly, it would be challenging to test this since the intrinsically faint radio emission would further suffer from scattering due to the very high absorption column density. Similarly, its $\simeq$6.5~kpc distance and location in the crowded Galactic center field, make it difficult to search for gamma-ray emission such as detected from \psr\ and \xss.

Support for this interpretation comes from the fact that the X-ray spectrum of \xmmbron, a power law with index $\Gamma\simeq1.4$, is harder than that of other neutron star LMXBs detected in the same luminosity range ($\Gamma\simeq1.7-2.7$; R. Wijnands et al. in preparation). This does not seem to be an effect of the high absorption ($N_{\mathrm{H}}\simeq10^{23}~\nh$), since other strongly absorbed neutron star LMXBs in the Galactic center display regular soft spectra  \citep[e.g., \ascabron, \grsbron, \ksbron;][]{maeda1996,sakano02,degenaar09_gc,degenaar2010_gc,degenaar2012_gc,degenaar2013_ks1741}. The M28 source also displays a strikingly hard X-ray spectrum ($\Gamma\simeq1.5$), and another common feature of the two is the lack of strong spectral evolution, despite 4--5 orders of magnitude change in X-ray luminosity \citep[][]{linares2014}. These similarities may provide additional support for hypothesis that the peculiar X-ray flux behavior of \xmmbron\ is related to the interaction between the accretion flow and the magnetic field of the neutron star.

\subsection{Similar Behavior in Other LMXBs?}
It is worth considering whether the X-ray behavior of \xmmbron\ (and the MRSP/LMXB transitional objects) is common. There are several neutron star LMXBs that have been observed at multiple epochs and appear to sustain a stable quiescent X-ray luminosity, or only show a steady decrease in thermal luminosity that is attributed to a gradual cooling of the hot neutron star \citep[e.g.,][]{cackett2008,cackett2010,degenaar2013_ter5,degenaar2014_exo3,servillat2012,guillot2013}. However, there are several other sources that display some form of flux variations in between their outburst ($L_{\mathrm{X}} \gtrsim 10^{36}~\lum$) and quiescence levels ($L_{\mathrm{X}} \lesssim 10^{33}~\lum$). 

Cen X-4, Aql X-1, \sax\ (a 2.5 ms X-ray pulsar), and \exoter\ (in the globular cluster Terzan 5) all vary at $L_{\mathrm{X}} \simeq10^{32}-10^{33}~\lum$ on a timescales of hours and months/years \citep[e.g.,][]{rutledge2002_aqlX1,campana2004,campana2008_saxj1808,cackett2010_cenx4,cackett2011,degenaar2012_1745}. Furthermore, \ksbron, \xte, \grsbron, and \saxrudy\ have shown ``mini'' accretion outbursts during which their luminosity increased from quiescence ($L_{\mathrm{X}} \simeq 10^{32}~\lum$) to $L_{\mathrm{X}} \simeq 10^{34}-10^{35}~\lum$ for a few days \citep[][]{degenaar2010_gc,degenaar2013_ks1741,fridriksson2011,wijnands2013}. Finally, \grobron\ (a 0.5-s X-ray pulsar) may also exhibit extended low-luminosity states at $L_{\mathrm{X}}\simeq10^{33}-10^{34}~\lum$ such as seen in \xmmbron\ \citep[e.g.,][]{wijnands2002_gro1744,degenaar2012_gc}, and somewhat similar quasi-stable low states are seen in 4U 1608--52 and \saxbron, albeit at a higher luminosity of $L_{\mathrm{X}}\simeq10^{35}~\lum$ \citep[e.g.,][]{wijnands2002_saxj1747,simon2004,degenaar2012_gc}. 

There is thus considerable X-ray variability observed for neutron star LMXBs in between their outburst and quiescent levels. Two of the sources mentioned above (\sax\ and \grobron) are known to be X-ray pulsars, indicating the magnetic field of these neutron stars could be involved in the X-ray variability. Indeed, it has been proposed that \sax\ acts as a MSRP during X-ray quiescent state, although no radio pulsations have been detected \citep[e.g.,][]{buderi2003}. It is unclear, however, if the interaction of the neutron star magnetic field and the accretion flow can account for the X-ray variability in all these LMXBs. For instance, LMXBs harboring a black hole instead of a neutron star also display considerably X-ray variability in between outburst and quiescence \citep[e.g.,][]{kong2002,hynes2004,tomsick2004_BH}, which clearly requires a different explanation.


\acknowledgments
N.D. is supported by NASA through Hubble Postdoctoral Fellowship grant number HST-HF-51287.01-A from the Space Telescope Science Institute, which is operated by the Association of Universities for Research in Astronomy, Incorporated, under NASA contract NAS5-26555. D.A. acknowledges support from the Royal Society. G.P. is supported by an EU Marie Curie Intra-European Fellowship under contract number EFP7-PEOPLE-2012-IEF-331095. This work made use of the public data archives of \swift, \chan\ and \xmm. 

{\it Facilities:} \facility{{\it Swift} (XRT), {\it XMM} (EPIC), {\it CXO} (HRC,ACIS)}




\begin{thebibliography}{96}
\expandafter\ifx\csname natexlab\endcsname\relax\def\natexlab#1{#1}\fi

\bibitem[{{Alpar} {et~al.}(1982){Alpar}, {Cheng}, {Ruderman}, \&
  {Shaham}}]{alpar1982}
{Alpar}, M.~A., {Cheng}, A.~F., {Ruderman}, M.~A., \& {Shaham}, J. 1982, \nat,
  300, 728

\bibitem[{{Archibald} {et~al.}(2009){Archibald}, {Stairs}, {Ransom}, {Kaspi},
  {Kondratiev}, {Lorimer}, {McLaughlin}, {Boyles}, {Hessels}, {Lynch}, {van
  Leeuwen}, {Roberts}, {Jenet}, {Champion}, {Rosen}, {Barlow}, {Dunlap}, \&
  {Remillard}}]{archibald2009}
{Archibald}, A.~M., {Stairs}, I.~H., {Ransom}, S.~M., {et~al.} 2009, Science,
  324, 1411

\bibitem[{{Arnaud}(1996)}]{xspec}
{Arnaud}, K.~A. 1996, in Astronomical Society of the Pacific Conference Series,
  Vol. 101, Astronomical Data Analysis Software and Systems V, ed. G.~H.
  {Jacoby} \& J.~{Barnes}, 17

\bibitem[{{Barlow} {et~al.}(2012){Barlow}, {Barthelmy}, {Gronwall}, {Palmer},
  \& {Zhang}}]{barlow2012}
{Barlow}, B.~N., {Barthelmy}, S.~D., {Gronwall}, C., {Palmer}, D.~M., \&
  {Zhang}, B.-B. 2012, GRB Coordinates Network, 13619

\bibitem[{{Barthelmy} {et~al.}(2005){Barthelmy}, {Barbier}, {Cummings},
  {Fenimore}, {Gehrels}, {Hullinger}, {Krimm}, {Markwardt}, {Palmer},
  {Parsons}, {Sato}, {Suzuki}, {Takahashi}, {Tashiro}, \&
  {Tueller}}]{barthelmy05}
{Barthelmy}, S.~D., {Barbier}, L.~M., {Cummings}, J.~R., {et~al.} 2005, Space
  Science Reviews, 120, 143

\bibitem[{{Bassa} {et~al.}(2014){Bassa}, {Patruno}, {Hessels}, {Keane},
  {Monard}, {Mahony}, {Bogdanov}, {Corbel}, {Edwards}, {Archibald}, {Janssen},
  {Stappers}, \& {Tendulkar}}]{bassa2014}
{Bassa}, C.~G., {Patruno}, A., {Hessels}, J.~W.~T., {et~al.} 2014, \mnras, 441,
  1825

\bibitem[{{Bhattacharya} \& {van den Heuvel}(1991)}]{bhattacharya1991}
{Bhattacharya}, D., \& {van den Heuvel}, E.~P.~J. 1991, \physrep, 203, 1

\bibitem[{{Bildsten}(1998)}]{bildsten98}
{Bildsten}, L. 1998, in NATO ASIC Proc. 515: The Many Faces of Neutron Stars.,
  ed. R.~{Buccheri}, J.~{van Paradijs}, \& A.~{Alpar}, 419

\bibitem[{{Bogdanov} {et~al.}(2011){Bogdanov}, {Archibald}, {Hessels}, {Kaspi},
  {Lorimer}, {McLaughlin}, {Ransom}, \& {Stairs}}]{bogdanov2011}
{Bogdanov}, S., {Archibald}, A.~M., {Hessels}, J.~W.~T., {et~al.} 2011, \apj,
  742, 97

\bibitem[{{Bogdanov} {et~al.}(2014){Bogdanov}, {Patruno}, {Archibald}, {Bassa},
  {Hessels}, {Janssen}, \& {Stappers}}]{bogdanov2014}
{Bogdanov}, S., {Patruno}, A., {Archibald}, A.~M., {et~al.} 2014, \apj, 789, 40

\bibitem[{{Burderi} {et~al.}(2003){Burderi}, {Di Salvo}, {D'Antona}, {Robba},
  \& {Testa}}]{buderi2003}
{Burderi}, L., {Di Salvo}, T., {D'Antona}, F., {Robba}, N.~R., \& {Testa}, V.
  2003, \aap, 404, L43

\bibitem[{{Burrows} {et~al.}(2005){Burrows}, {Hill}, {Nousek}, {Kennea},
  {Wells}, {Osborne}, {Abbey}, {Beardmore}, {Mukerjee}, {Short}, {Chincarini},
  {Campana}, {Citterio}, {Moretti}, {Pagani}, {Tagliaferri}, {Giommi},
  {Capalbi}, {Tamburelli}, {Angelini}, {Cusumano}, {Br{\"a}uninger}, {Burkert},
  \& {Hartner}}]{burrows05}
{Burrows}, D.~N., {Hill}, J.~E., {Nousek}, J.~A., {et~al.} 2005, Space Science
  Reviews, 120, 165

\bibitem[{{Cackett} {et~al.}(2010{\natexlab{a}}){Cackett}, {Brown}, {Cumming},
  {Degenaar}, {Miller}, \& {Wijnands}}]{cackett2010}
{Cackett}, E.~M., {Brown}, E.~F., {Cumming}, A., {et~al.} 2010{\natexlab{a}},
  \apjl, 722, L137

\bibitem[{{Cackett} {et~al.}(2010{\natexlab{b}}){Cackett}, {Brown}, {Miller},
  \& {Wijnands}}]{cackett2010_cenx4}
{Cackett}, E.~M., {Brown}, E.~F., {Miller}, J.~M., \& {Wijnands}, R.
  2010{\natexlab{b}}, \apj, 720, 1325

\bibitem[{{Cackett} {et~al.}(2011){Cackett}, {Fridriksson}, {Homan}, {Miller},
  \& {Wijnands}}]{cackett2011}
{Cackett}, E.~M., {Fridriksson}, J.~K., {Homan}, J., {Miller}, J.~M., \&
  {Wijnands}, R. 2011, \mnras, 414, 3006

\bibitem[{{Cackett} {et~al.}(2008){Cackett}, {Wijnands}, {Miller}, {Brown}, \&
  {Degenaar}}]{cackett2008}
{Cackett}, E.~M., {Wijnands}, R., {Miller}, J.~M., {Brown}, E.~F., \&
  {Degenaar}, N. 2008, \apjl, 687, L87

\bibitem[{{Campana} {et~al.}(2004){Campana}, {Israel}, {Stella}, {Gastaldello},
  \& {Mereghetti}}]{campana2004}
{Campana}, S., {Israel}, G.~L., {Stella}, L., {Gastaldello}, F., \&
  {Mereghetti}, S. 2004, \apj, 601, 474

\bibitem[{{Campana} {et~al.}(2008){Campana}, {Stella}, \&
  {Kennea}}]{campana2008_saxj1808}
{Campana}, S., {Stella}, L., \& {Kennea}, J.~A. 2008, \apjl, 684, L99

\bibitem[{{Chen} {et~al.}(1997){Chen}, {Shrader}, \& {Livio}}]{chen97}
{Chen}, W., {Shrader}, C.~R., \& {Livio}, M. 1997, \apj, 491, 312

\bibitem[{{Coriat} {et~al.}(2012){Coriat}, {Fender}, \&
  {Dubus}}]{coriat2012_dim}
{Coriat}, M., {Fender}, R.~P., \& {Dubus}, G. 2012, \mnras, 424, 1991

\bibitem[{{de Martino} {et~al.}(2013){de Martino}, {Belloni}, {Falanga},
  {Papitto}, {Motta}, {Pellizzoni}, {Evangelista}, {Piano}, {Masetti},
  {Bonnet-Bidaud}, {Mouchet}, {Mukai}, \& {Possenti}}]{demartino2013}
{de Martino}, D., {Belloni}, T., {Falanga}, M., {et~al.} 2013, \aap, 550, A89

\bibitem[{{Degenaar} {et~al.}(2012{\natexlab{a}}){Degenaar}, {Kennea},
  {Wijnands}, {Miller}, \& {Gehrels}}]{degenaar2012_xmmburst}
{Degenaar}, N., {Kennea}, J.~A., {Wijnands}, R., {Miller}, J.~M., \& {Gehrels},
  N. 2012{\natexlab{a}}, \atel, 4308

\bibitem[{{Degenaar} {et~al.}(2013{\natexlab{a}}){Degenaar}, {Miller},
  {Wijnands}, {Altamirano}, \& {Fabian}}]{degenaar2013_igrj1706}
{Degenaar}, N., {Miller}, J.~M., {Wijnands}, R., {Altamirano}, D., \& {Fabian},
  A.~C. 2013{\natexlab{a}}, \apjl, 767, L37

\bibitem[{{Degenaar} \& {Wijnands}(2009)}]{degenaar09_gc}
{Degenaar}, N., \& {Wijnands}, R. 2009, \aap, 495, 547

\bibitem[{{Degenaar} \& {Wijnands}(2010)}]{degenaar2010_gc}
---. 2010, \aap, 524, A69

\bibitem[{{Degenaar} \& {Wijnands}(2012)}]{degenaar2012_1745}
---. 2012, \mnras, 422, 581

\bibitem[{{Degenaar} \& {Wijnands}(2013)}]{degenaar2013_ks1741}
{Degenaar}, N., \& {Wijnands}, R. 2013, in IAU Symposium, Vol. 290, IAU
  Symposium, ed. C.~M. {Zhang}, T.~{Belloni}, M.~{M{\'e}ndez}, \& S.~N.
  {Zhang}, 113--116

\bibitem[{{Degenaar} {et~al.}(2012{\natexlab{b}}){Degenaar}, {Wijnands},
  {Cackett}, {Homan}, {in 't Zand}, {Kuulkers}, {Maccarone}, \& {van der
  Klis}}]{degenaar2012_gc}
{Degenaar}, N., {Wijnands}, R., {Cackett}, E.~M., {et~al.} 2012{\natexlab{b}},
  \aap, 545, A49

\bibitem[{{Degenaar} {et~al.}(2011){Degenaar}, {Wijnands}, \&
  {Kaur}}]{degenaar2011_burst}
{Degenaar}, N., {Wijnands}, R., \& {Kaur}, R. 2011, \mnras, 414, L104

\bibitem[{{Degenaar} {et~al.}(2010{\natexlab{a}}){Degenaar}, {Wijnands},
  {Kennea}, \& {Gehrels}}]{degenaar2010_atel_grs_xmm}
{Degenaar}, N., {Wijnands}, R., {Kennea}, J., \& {Gehrels}, N.
  2010{\natexlab{a}}, \atel, 2770

\bibitem[{{Degenaar} {et~al.}(2012{\natexlab{c}}){Degenaar}, {Wijnands},
  {Kennea}, {Miller}, \& {Gehrels}}]{degenaar2012_xmmactive}
{Degenaar}, N., {Wijnands}, R., {Kennea}, J.~A., {Miller}, J.~M., \& {Gehrels},
  N. 2012{\natexlab{c}}, \atel, 4305

\bibitem[{{Degenaar} {et~al.}(2010{\natexlab{b}}){Degenaar}, {Jonker},
  {Torres}, {Kaur}, {Rea}, {Israel}, {Patruno}, {Trap}, {Cackett}, {D'Avanzo},
  {Lo Curto}, {Novara}, {Krimm}, {Holland}, {de Luca}, {Esposito}, \&
  {Wijnands}}]{degenaar2010_burst}
{Degenaar}, N., {Jonker}, P.~G., {Torres}, M.~A.~P., {et~al.}
  2010{\natexlab{b}}, \mnras, 404, 1591

\bibitem[{{Degenaar} {et~al.}(2013{\natexlab{b}}){Degenaar}, {Wijnands},
  {Brown}, {Altamirano}, {Cackett}, {Fridriksson}, {Homan}, {Heinke}, {Miller},
  {Pooley}, \& {Sivakoff}}]{degenaar2013_ter5}
{Degenaar}, N., {Wijnands}, R., {Brown}, E.~F., {et~al.} 2013{\natexlab{b}},
  \apj, 775, 48

\bibitem[{{Degenaar} {et~al.}(2014){Degenaar}, {Medin}, {Cumming}, {Wijnands},
  {Wolff}, {Cackett}, {Miller}, {Jonker}, {Homan}, \&
  {Brown}}]{degenaar2014_exo3}
{Degenaar}, N., {Medin}, Z., {Cumming}, A., {et~al.} 2014, ArXiv:1403.2385

\bibitem[{{Drave} {et~al.}(2014){Drave}, {Bird}, {Sidoli}, {Sguera}, {Bazzano},
  {Hill}, \& {Goossens}}]{drave2014}
{Drave}, S.~P., {Bird}, A.~J., {Sidoli}, L., {et~al.} 2014, \mnras, 439, 2175

\bibitem[{{Ek{\c s}{\.I}} \& {Alpar}(2005)}]{eksi2005}
{Ek{\c s}{\.I}}, K.~Y., \& {Alpar}, M.~A. 2005, \apj, 620, 390

\bibitem[{{Enoto} {et~al.}(2014){Enoto}, {Sasano}, {Yamada}, {Tamagawa},
  {Makishima}, {Pottschmidt}, {Marcu}, {Corbet}, {Fuerst}, \&
  {Wilms}}]{enoto2014}
{Enoto}, T., {Sasano}, M., {Yamada}, S., {et~al.} 2014, \apj, 786, 127

\bibitem[{{Falanga} {et~al.}(2008){Falanga}, {Chenevez}, {Cumming}, {Kuulkers},
  {Trap}, \& {Goldwurm}}]{falanga08}
{Falanga}, M., {Chenevez}, J., {Cumming}, A., {et~al.} 2008, \aap, 484, 43

\bibitem[{{Fridriksson} {et~al.}(2011){Fridriksson}, {Homan}, {Wijnands},
  {Cackett}, {Altamirano}, {Degenaar}, {Brown}, {M{\'e}ndez}, \&
  {Belloni}}]{fridriksson2011}
{Fridriksson}, J.~K., {Homan}, J., {Wijnands}, R., {et~al.} 2011, \apj, 736,
  162

\bibitem[{{Fujimoto} {et~al.}(1981){Fujimoto}, {Hanawa}, \&
  {Miyaji}}]{fujimoto81}
{Fujimoto}, M.~Y., {Hanawa}, T., \& {Miyaji}, S. 1981, \apj, 247, 267

\bibitem[{{Galloway} {et~al.}(2008){Galloway}, {Muno}, {Hartman}, {Psaltis}, \&
  {Chakrabarty}}]{galloway06}
{Galloway}, D.~K., {Muno}, M.~P., {Hartman}, J.~M., {Psaltis}, D., \&
  {Chakrabarty}, D. 2008, \apjs, 179, 360

\bibitem[{{Garmire} {et~al.}(2003){Garmire}, {Bautz}, {Ford}, {Nousek}, \&
  {Ricker}}]{garmire2003_acis}
{Garmire}, G.~P., {Bautz}, M.~W., {Ford}, P.~G., {Nousek}, J.~A., \& {Ricker},
  Jr., G.~R. 2003, in Presented at the Society of Photo-Optical Instrumentation
  Engineers (SPIE) Conference, Vol. 4851, Society of Photo-Optical
  Instrumentation Engineers (SPIE) Conference Series, ed. {J.~E.~Truemper \&
  H.~D.~Tananbaum}, 28

\bibitem[{{Guillot} {et~al.}(2013){Guillot}, {Servillat}, {Webb}, \&
  {Rutledge}}]{guillot2013}
{Guillot}, S., {Servillat}, M., {Webb}, N.~A., \& {Rutledge}, R.~E. 2013, \apj,
  772, 7

\bibitem[{{Hameury} \& {Lasota}(2014)}]{hameury2014}
{Hameury}, J.-M., \& {Lasota}, J.-P. 2014, ArXiv:1407.3156

\bibitem[{{Hill} {et~al.}(2011){Hill}, {Szostek}, {Corbel}, {Camilo}, {Corbet},
  {Dubois}, {Dubus}, {Edwards}, {Ferrara}, {Kerr}, {Koerding}, {Kozie{\l}}, \&
  {Stawarz}}]{hill2011}
{Hill}, A.~B., {Szostek}, A., {Corbel}, S., {et~al.} 2011, \mnras, 415, 235

\bibitem[{{Hill} {et~al.}(2004){Hill}, {Burrows}, {Nousek}, {Abbey}, {Ambrosi},
  {Br{\"a}uninger}, {Burkert}, {Campana}, {Cheruvu}, {Cusumano}, {Freyberg},
  {Hartner}, {Klar}, {Mangels}, {Moretti}, {Mori}, {Morris}, {Short},
  {Tagliaferri}, {Watson}, {Wood}, \& {Wells}}]{hill2004}
{Hill}, J.~E., {Burrows}, D.~N., {Nousek}, J.~A., {et~al.} 2004, in Society of
  Photo-Optical Instrumentation Engineers (SPIE) Conference Series, Vol. 5165,
  Society of Photo-Optical Instrumentation Engineers (SPIE) Conference Series,
  ed. K.~A. {Flanagan} \& O.~H.~W. {Siegmund}, 217--231

\bibitem[{{Hynes} {et~al.}(2004){Hynes}, {Charles}, {Garcia}, {Robinson},
  {Casares}, {Haswell}, {Kong}, {Rupen}, {Fender}, {Wagner}, {Gallo}, {Eves},
  {Shahbaz}, \& {Zurita}}]{hynes2004}
{Hynes}, R.~I., {Charles}, P.~A., {Garcia}, M.~R., {et~al.} 2004, \apjl, 611,
  L125

\bibitem[{{in 't Zand} {et~al.}(2005){in 't Zand}, {Cornelisse}, \&
  {M{\'e}ndez}}]{zand05}
{in 't Zand}, J.~J.~M., {Cornelisse}, R., \& {M{\'e}ndez}, M. 2005, \aap, 440,
  287

\bibitem[{{in 't Zand} {et~al.}(2007){in 't Zand}, {Jonker}, \&
  {Markwardt}}]{zand07}
{in 't Zand}, J.~J.~M., {Jonker}, P.~G., \& {Markwardt}, C.~B. 2007, \aap, 465,
  953

\bibitem[{{Kenter} {et~al.}(2000){Kenter}, {Chappell}, {Kraft}, {Meehan},
  {Murray}, {Zombeck}, {Hole}, {Juda}, {Donnelly}, {Patnaude}, {Pease},
  {Wilton}, {Zhao}, {Austin}, {Fraser}, {Pearson}, {Lees}, {Brunton},
  {Barbera}, {Collura}, \& {Serio}}]{kenter2000_hrc}
{Kenter}, A.~T., {Chappell}, J.~H., {Kraft}, R.~P., {et~al.} 2000, in Presented
  at the Society of Photo-Optical Instrumentation Engineers (SPIE) Conference,
  Vol. 4012, Society of Photo-Optical Instrumentation Engineers (SPIE)
  Conference Series, ed. {J.~E.~Truemper \& B.~Aschenbach}, 467

\bibitem[{{King} {et~al.}(1996){King}, {Kolb}, \& {Burderi}}]{king1996}
{King}, A.~R., {Kolb}, U., \& {Burderi}, L. 1996, \apjl, 464, L127

\bibitem[{{Kong} {et~al.}(2002){Kong}, {McClintock}, {Garcia}, {Murray}, \&
  {Barret}}]{kong2002}
{Kong}, A.~K.~H., {McClintock}, J.~E., {Garcia}, M.~R., {Murray}, S.~S., \&
  {Barret}, D. 2002, \apj, 570, 277

\bibitem[{{Kuulkers} {et~al.}(2003){Kuulkers}, {den Hartog}, {in 't Zand},
  {Verbunt}, {Harris}, \& {Cocchi}}]{kuulkers2003}
{Kuulkers}, E., {den Hartog}, P.~R., {in 't Zand}, J.~J.~M., {et~al.} 2003,
  \aap, 399, 663

\bibitem[{{Lasota}(2001)}]{lasota01}
{Lasota}, J.-P. 2001, New Astronomy Review, 45, 449

\bibitem[{{Linares}(2014)}]{linares2014_redbacks}
{Linares}, M. 2014, ArXiv:1406.2384

\bibitem[{{Linares} {et~al.}(2009){Linares}, {Watts}, {Wijnands}, {Soleri},
  {Degenaar}, {Curran}, {Starling}, \& {van der Klis}}]{linares09}
{Linares}, M., {Watts}, A.~L., {Wijnands}, R., {et~al.} 2009, \mnras, 392, L11

\bibitem[{{Linares} {et~al.}(2014){Linares}, {Bahramian}, {Heinke}, {Wijnands},
  {Patruno}, {Altamirano}, {Homan}, {Bogdanov}, \& {Pooley}}]{linares2014}
{Linares}, M., {Bahramian}, A., {Heinke}, C., {et~al.} 2014, \mnras, 438, 251

\bibitem[{{Lipunov} {et~al.}(1992){Lipunov}, {B{\"o}rner}, \&
  {Wadhwa}}]{lipunov1992}
{Lipunov}, V.~M., {B{\"o}rner}, G., \& {Wadhwa}, R.~S. 1992, {Astrophysics of
  Neutron Stars}

\bibitem[{{Maccarone} \& {Patruno}(2013)}]{maccarone2013}
{Maccarone}, T.~J., \& {Patruno}, A. 2013, \mnras, 428, 1335

\bibitem[{{Maeda} {et~al.}(1996){Maeda}, {Koyama}, {Sakano}, {Takeshima}, \&
  {Yamauchi}}]{maeda1996}
{Maeda}, Y., {Koyama}, K., {Sakano}, M., {Takeshima}, T., \& {Yamauchi}, S.
  1996, \pasj, 48, 417

\bibitem[{{Masetti} {et~al.}(2007){Masetti}, {Landi}, {Pretorius}, {Sguera},
  {Bird}, {Perri}, {Charles}, {Kennea}, {Malizia}, \& {Ubertini}}]{masetti2007}
{Masetti}, N., {Landi}, R., {Pretorius}, M.~L., {et~al.} 2007, \aap, 470, 331

\bibitem[{{Mauerhan} {et~al.}(2009){Mauerhan}, {Muno}, {Morris}, {Bauer},
  {Nishiyama}, \& {Nagata}}]{mauerhan2009}
{Mauerhan}, J.~C., {Muno}, M.~P., {Morris}, M.~R., {et~al.} 2009, \apj, 703, 30

\bibitem[{{Meyer} \& {Meyer-Hofmeister}(1983)}]{meyer1983}
{Meyer}, F., \& {Meyer-Hofmeister}, E. 1983, \aap, 121, 29

\bibitem[{{Papitto} {et~al.}(2013{\natexlab{a}}){Papitto}, {Bozzo}, {Ferrigno},
  {Pavan}, {Romano}, \& {Campana}}]{papitto2013}
{Papitto}, A., {Bozzo}, E., {Ferrigno}, C., {et~al.} 2013{\natexlab{a}}, \atel,
  4959

\bibitem[{{Papitto} {et~al.}(2014){Papitto}, {Torres}, \& {Li}}]{papitto2014}
{Papitto}, A., {Torres}, D.~F., \& {Li}, J. 2014, \mnras, 438, 2105

\bibitem[{{Papitto} {et~al.}(2013{\natexlab{b}}){Papitto}, {Ferrigno}, {Bozzo},
  {Rea}, {Pavan}, {Burderi}, {Burgay}, {Campana}, {di Salvo}, {Falanga},
  {Filipovi{\'c}}, {Freire}, {Hessels}, {Possenti}, {Ransom}, {Riggio},
  {Romano}, {Sarkissian}, {Stairs}, {Stella}, {Torres}, {Wieringa}, \&
  {Wong}}]{papitto2013_nature}
{Papitto}, A., {Ferrigno}, C., {Bozzo}, E., {et~al.} 2013{\natexlab{b}}, \nat,
  501, 517

\bibitem[{{Papitto} {et~al.}(2013{\natexlab{c}}){Papitto}, {Hessels}, {Burgay},
  {Ransom}, {Rea}, {Possenti}, {Stairs}, {Ferrigno}, \&
  {Bozz}}]{papitto2013_radio}
{Papitto}, A., {Hessels}, J.~W.~T., {Burgay}, M., {et~al.} 2013{\natexlab{c}},
  \atel, 5069

\bibitem[{{Patel} {et~al.}(2007){Patel}, {Zurita}, {Del Santo}, {Finger},
  {Kouveliotou}, {Eichler}, {G{\"o}{\v g}{\"u}{\c s}}, {Ubertini}, {Walter},
  {Woods}, {Wilson}, {Wachter}, \& {Bazzano}}]{patel2007}
{Patel}, S.~K., {Zurita}, J., {Del Santo}, M., {et~al.} 2007, \apj, 657, 994

\bibitem[{{Patruno} {et~al.}(2014){Patruno}, {Archibald}, {Hessels},
  {Bogdanov}, {Stappers}, {Bassa}, {Janssen}, {Kaspi}, {Tendulkar}, \&
  {Lyne}}]{patruno2014}
{Patruno}, A., {Archibald}, A.~M., {Hessels}, J.~W.~T., {et~al.} 2014, \apjl,
  781, L3

\bibitem[{{Peng} {et~al.}(2007){Peng}, {Brown}, \& {Truran}}]{peng2007}
{Peng}, F., {Brown}, E.~F., \& {Truran}, J.~W. 2007, \apj, 654, 1022

\bibitem[{{Romano} {et~al.}(2011){Romano}, {La Parola}, {Vercellone},
  {Cusumano}, {Sidoli}, {Krimm}, {Pagani}, {Esposito}, {Hoversten}, {Kennea},
  {Page}, {Burrows}, \& {Gehrels}}]{romano2011}
{Romano}, P., {La Parola}, V., {Vercellone}, S., {et~al.} 2011, \mnras, 410,
  1825

\bibitem[{{Roy} {et~al.}(2014){Roy}, {Bhattacharyya}, \& {Ray}}]{roy2014}
{Roy}, J., {Bhattacharyya}, B., \& {Ray}, P.~S. 2014, \atel, 5890

\bibitem[{{Rutledge} {et~al.}(2002){Rutledge}, {Bildsten}, {Brown}, {Pavlov},
  \& {Zavlin}}]{rutledge2002_aqlX1}
{Rutledge}, R.~E., {Bildsten}, L., {Brown}, E.~F., {Pavlov}, G.~G., \&
  {Zavlin}, V.~E. 2002, \apj, 577, 346

\bibitem[{{Sakano} {et~al.}(2002){Sakano}, {Koyama}, {Murakami}, {Maeda}, \&
  {Yamauchi}}]{sakano02}
{Sakano}, M., {Koyama}, K., {Murakami}, H., {Maeda}, Y., \& {Yamauchi}, S.
  2002, \apjs, 138, 19

\bibitem[{{Sakano} {et~al.}(2005){Sakano}, {Warwick}, {Decourchelle}, \&
  {Wang}}]{sakano05}
{Sakano}, M., {Warwick}, R.~S., {Decourchelle}, A., \& {Wang}, Q.~D. 2005,
  \mnras, 357, 1211

\bibitem[{{Servillat} {et~al.}(2012){Servillat}, {Heinke}, {Ho}, {Grindlay},
  {Hong}, {van den Berg}, \& {Bogdanov}}]{servillat2012}
{Servillat}, M., {Heinke}, C.~O., {Ho}, W.~C.~G., {et~al.} 2012, \mnras, 423,
  1556

\bibitem[{{Sidoli}(2013)}]{sidoli2013}
{Sidoli}, L. 2013, ArXiv:1301.7574

\bibitem[{{Simonsen}(2011)}]{simonsen2011}
{Simonsen}, M. 2011, Journal of the American Association of Variable Star
  Observers (JAAVSO), 39, 66

\bibitem[{{Simonsen} {et~al.}(2014){Simonsen}, {Bohlsen}, {Hambsch}, \&
  {Stubbings}}]{simonsen2014}
{Simonsen}, M., {Bohlsen}, T., {Hambsch}, F.-J., \& {Stubbings}, R. 2014,
  Journal of the American Association of Variable Star Observers (JAAVSO), 42,
  199

\bibitem[{{Stappers} {et~al.}(2014){Stappers}, {Archibald}, {Hessels}, {Bassa},
  {Bogdanov}, {Janssen}, {Kaspi}, {Lyne}, {Patruno}, {Tendulkar}, {Hill}, \&
  {Glanzman}}]{stappers2013}
{Stappers}, B.~W., {Archibald}, A.~M., {Hessels}, J.~W.~T., {et~al.} 2014,
  \apj, 790, 39

\bibitem[{{Strohmayer} \& {Bildsten}(2006)}]{strohmayer06}
{Strohmayer}, T., \& {Bildsten}, L. 2006, {New views of thermonuclear bursts},
  ed. M.~Lewin, W. H. G. \& van der~Klis, 113--156

\bibitem[{{Strohmayer} {et~al.}(1996){Strohmayer}, {Zhang}, {Swank}, {Smale},
  {Titarchuk}, {Day}, \& {Lee}}]{strohmayer1996}
{Strohmayer}, T.~E., {Zhang}, W., {Swank}, J.~H., {et~al.} 1996, \apjl, 469, L9

\bibitem[{{Str{\"u}der} {et~al.}(2001){Str{\"u}der}, {Briel}, {Dennerl},
  {Hartmann}, {Kendziorra}, {Meidinger}, {Pfeffermann}, {Reppin}, {Aschenbach},
  {Bornemann}, {Br{\"a}uninger}, {Burkert}, {Elender}, {Freyberg}, {Haberl},
  {Hartner}, {Heuschmann}, {Hippmann}, {Kastelic}, {Kemmer}, {Kettenring},
  {Kink}, {Krause}, {M{\"u}ller}, {Oppitz}, {Pietsch}, {Popp}, {Predehl},
  {Read}, {Stephan}, {St{\"o}tter}, {Tr{\"u}mper}, {Holl}, {Kemmer}, {Soltau},
  {St{\"o}tter}, {Weber}, {Weichert}, {von Zanthier}, {Carathanassis}, {Lutz},
  {Richter}, {Solc}, {B{\"o}ttcher}, {Kuster}, {Staubert}, {Abbey}, {Holland},
  {Turner}, {Balasini}, {Bignami}, {La Palombara}, {Villa}, {Buttler},
  {Gianini}, {Lain{\'e}}, {Lumb}, \& {Dhez}}]{struder2001_pn}
{Str{\"u}der}, L., {Briel}, U., {Dennerl}, K., {et~al.} 2001, \aap, 365, L18

\bibitem[{{Szkody} {et~al.}(2013){Szkody}, {Albright}, {Linnell}, {Everett},
  {McMillan}, {Saurage}, {Huehnerhoff}, {Howell}, {Simonsen}, \&
  {Hunt-Walker}}]{szkody2013}
{Szkody}, P., {Albright}, M., {Linnell}, A.~P., {et~al.} 2013, \pasp, 125, 1421

\bibitem[{{Tauris}(2012)}]{tauris2012}
{Tauris}, T.~M. 2012, Science, 335, 561

\bibitem[{{Tomsick} {et~al.}(2004){Tomsick}, {Kalemci}, \&
  {Kaaret}}]{tomsick2004_BH}
{Tomsick}, J.~A., {Kalemci}, E., \& {Kaaret}, P. 2004, \apj, 601, 439

\bibitem[{{Turner} {et~al.}(2001){Turner}, {Abbey}, {Arnaud}, {Balasini},
  {Barbera}, {Belsole}, {Bennie}, {Bernard}, {Bignami}, {Boer}, {Briel},
  {Butler}, {Cara}, {Chabaud}, {Cole}, {Collura}, {Conte}, {Cros}, {Denby},
  {Dhez}, {Di Coco}, {Dowson}, {Ferrando}, {Ghizzardi}, {Gianotti}, {Goodall},
  {Gretton}, {Griffiths}, {Hainaut}, {Hochedez}, {Holland}, {Jourdain},
  {Kendziorra}, {Lagostina}, {Laine}, {La Palombara}, {Lortholary}, {Lumb},
  {Marty}, {Molendi}, {Pigot}, {Poindron}, {Pounds}, {Reeves}, {Reppin},
  {Rothenflug}, {Salvetat}, {Sauvageot}, {Schmitt}, {Sembay}, {Short},
  {Spragg}, {Stephen}, {Str{\"u}der}, {Tiengo}, {Trifoglio}, {Tr{\"u}mper},
  {Vercellone}, {Vigroux}, {Villa}, {Ward}, {Whitehead}, \&
  {Zonca}}]{turner2001_mos}
{Turner}, M.~J.~L., {Abbey}, A., {Arnaud}, M., {et~al.} 2001, \aap, 365, L27

\bibitem[{{{\v S}imon}(2004)}]{simon2004}
{{\v S}imon}, V. 2004, \aap, 418, 617

\bibitem[{{van Paradijs}(1996)}]{vanparadijs1996}
{van Paradijs}, J. 1996, \apjl, 464, L139

\bibitem[{{Verner} {et~al.}(1996){Verner}, {Ferland}, {Korista}, \&
  {Yakovlev}}]{verner1996}
{Verner}, D.~A., {Ferland}, G.~J., {Korista}, K.~T., \& {Yakovlev}, D.~G. 1996,
  \apj, 465, 487

\bibitem[{{Wijnands} \& {Degenaar}(2013)}]{wijnands2013}
{Wijnands}, R., \& {Degenaar}, N. 2013, \mnras, 434, 1599

\bibitem[{{Wijnands} {et~al.}(2002){Wijnands}, {Miller}, \&
  {Wang}}]{wijnands2002_saxj1747}
{Wijnands}, R., {Miller}, J.~M., \& {Wang}, Q.~D. 2002, \apj, 579, 422

\bibitem[{{Wijnands} \& {van der Klis}(1998)}]{wijnands1998}
{Wijnands}, R., \& {van der Klis}, M. 1998, \nat, 394, 344

\bibitem[{{Wijnands} \& {Wang}(2002)}]{wijnands2002_gro1744}
{Wijnands}, R., \& {Wang}, Q.~D. 2002, \apjl, 568, L93

\bibitem[{{Wijnands} {et~al.}(2006){Wijnands}, {in 't Zand}, {Rupen},
  {Maccarone}, {Homan}, {Cornelisse}, {Fender}, {Grindlay}, {van der Klis},
  {Kuulkers}, {Markwardt}, {Miller-Jones}, \& {Wang}}]{wijnands06}
{Wijnands}, R., {in 't Zand}, J.~J.~M., {Rupen}, M., {et~al.} 2006, \aap, 449,
  1117

\bibitem[{{Wilms} {et~al.}(2000){Wilms}, {Allen}, \& {McCray}}]{wilms2000}
{Wilms}, J., {Allen}, A., \& {McCray}, R. 2000, \apj, 542, 914

\end{thebibliography}
\end{document}